\documentclass[aps,reprint,twocolumn,pre,10pt]{revtex4-1}
\usepackage[english]{babel}
\usepackage[latin1]{inputenc}
\usepackage{graphicx}
\usepackage{amsmath}
\usepackage{color}
\usepackage{tikz}
\usepackage{bm}
\addto\captionsenglish{}

\newcommand{\E}[1]{\left\langle #1\right\rangle}

\newcommand{\df}[2]{\frac{\partial #1}{\partial #2}}
\newcommand{\s}{\sigma}
\newcommand{\dt}{{\Delta t}}
\newcommand{\Info}{\mathcal{I}}
\renewcommand{\r}{\mathbf{r}}
\newcommand{\dr}{{\Delta\mathbf{r}}}
\newcommand{\nm}{\mathrm{nm}}
\newcommand{\mum}{\mu\mathrm{m}}

\newcommand{\SNR}{\kappa}
\newcommand{\CRB}{\mathrm{CRB}}
\newcommand{\ms}{\mathrm{ms}}
\renewcommand{\sec}{\mathrm{s}}
\newcommand{\ttot}{{t_{\rm tot}}}
\newcommand{\Ptot}{{P_{\rm tot}}}
\newcommand{\Pmin}{{P_{\rm min}}}

\newcommand{\dx}{{(\delta x)^2}}
\newcommand{\SE}[1]{{\rm SE}\left(#1\right)}

\begin{document}
\title{Optimizing experimental parameters for tracking of diffusing particles}
\date{\today}
\author{Christian L. Vestergaard}
\email{cvestergaard@gmail.com}
\affiliation{Department of Micro- and Nanotechnology, Technical University of Denmark, Kgs. Lyngby, DK-2800, Denmark, and Aix Marseille Universit\'{e}, Universit\'e de Toulon, CNRS, CPT, UMR 7332, 13288 Marseille, France (current address).}
\begin{abstract}
We describe how a single-particle tracking experiment should be designed in order for its recorded trajectories to contain the most information about a tracked particle's diffusion coefficient.
The precision of estimators for the diffusion coefficient is affected by motion blur, limited photon statistics, and the length of recorded time-series.
We demonstrate for a particle undergoing free diffusion that precision is negligibly affected by motion blur in typical experiments, while optimizing photon counts and the number of recorded frames is the key to precision.
Building on these results, we describe for a wide range of experimental scenarios how to choose experimental parameters in order to optimize the precision.
Generally, one should choose quantity over quality: experiments should be designed to maximize the number of frames recorded in a time-series, even if this means lower information content in individual frames.
\end{abstract}
\maketitle

\section{Introduction}
Single-particle tracking using time-lapse photography~\cite{Serge2008,Chenouard2014} enables investigation of diffusion of single molecules, e.g., proteins on cellular structures such as DNA~\cite{Tafvizi2011} and microtubules~\cite{Helenius2006}, on cell membranes~\cite{Serge2008,Wieser2008d}, and inside cells~\cite{Elf2007,Smith2011}.
Diffusion is ubiquitous at the microscopic level and precise determination of diffusion coefficients is paramount for understanding many chemical and biological processes.
Typical single-particle-tracking experiments consist in recording the photons emitted by a fluorescent particle (a fluorophore) using time-lapse photography, and determining the particle's positions from recorded images using a super-resolution localization technique~\cite{Serge2008,Mortensen2010,Deschout2012,Chenouard2014}, e.g., by fitting a Gaussian to the intensity profile in each recorded image.
The number of photons emitted by a fluorophore is limited, and traditionally, tracked particles have been recorded by leaving the camera shutter open continuously to maximize the number of photons recorded by the camera.
The time the camera's shutter stays open to take a single image, its {\sl exposure time}, is then equal to the time elapsed between consecutive images, the {\sl time-lapse} of recordings.
The motion of the tracked particle during the exposure time results in motion blur in the pictures (also referred to as {\sl dynamic error}), while diffraction and limited photon statistics result in localization error (also referred to as {\sl static error})~\cite{Savin2005,Berglund2010,Deschout2012}.
Additionally, the length of a time-series, i.e., the number of recorded positions, is usually limited, either due to bleaching of the fluorophore or due to the tracked particle diffusing out of the field-of-view.
All of the above adversely affect the precision of estimates of diffusion coefficients and make it important to get the most out of experimental data.

\begin{figure*}
  \includegraphics{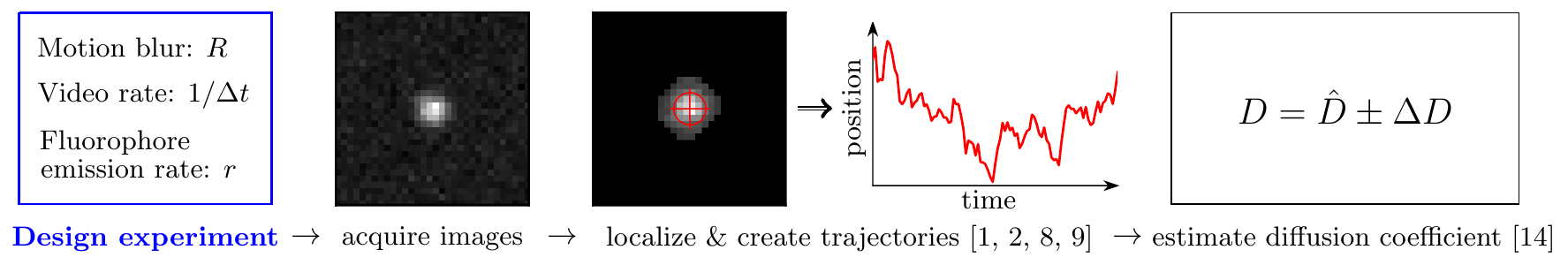}
  \caption{(Color online) Workflow for estimating diffusion coefficients from camera-based single-particle-tracking experiments.
  We are here concerned with optimizing the first step: how to choose experimental parameters (motion blur, video rate, and photon emission rate of the tracked fluorophore) for optimal precision of estimates of the diffusion coefficient of a tracked particle.
  The motion blur is characterized by a {\sl motion blur coefficient}, $R\in(0,1/4)$ [Eq.~(\ref{eq:R})]~\cite{Berglund2010}.
  The motion blur coefficient, $R$, and video rate---given by $1/\dt$, where $\dt$ is the time-lapse between measurements---may normally be controlled directly; the photon emission rate of the fluorophore, $r$, may be controlled indirectly, e.g., by varying the laser intensity or by the choice of fluorophore itself.
  Optimization of other steps in the workflow is addressed in the references given in the legends.}
  \label{fig:workflow}
\end{figure*}

A typical experiment for tracking single diffusing particles can be divided into multiple steps (Fig.~\ref{fig:workflow}):
(i) designing the experiment, e.g., choice of fluorophore and labeling technique, setting the video rate of the camera and the intensity of the illumination laser;
(ii) carrying out the experiment, i.e., recording images of the fluorescent particles;
(iii) treating images, localizing particles, and creating time-series of positions;
(iv) estimating diffusion coefficients from the time-series.
Optimal estimates of the particles' diffusion coefficients is obtained by optimizing each individual step.
Recently, the questions of how to best localize and track single particles~\cite{Mortensen2010,Deschout2012,Chenouard2014} and of how to optimally estimate diffusion coefficients from the resulting time-lapse-recorded trajectories~\cite{Berglund2010,Michalet2010,Michalet2012,Vestergaard2014} have been addressed.

With (near) optimal localization methods and estimators of diffusion coefficients at hand, we can now address the first step in the workflow: how should experiments be designed in order for recorded trajectories to contain the most information about diffusion coefficients?
One may turn several dials to influence the amount of information available for estimation of diffusion coefficients:
One may adjust both the video rate of the camera and the photon emission rate of the tracked fluorophores.
Furthermore, the motion blur in recorded images can be controlled by leaving the shutter open for only part of the time-lapse, following a given shutter sequence.
The advent of stroboscopic tracking techniques~\cite{Elf2007}, which synchronize illumination and recording of the sample, makes it possible to control the motion blur without sacrificing photon economy.

Recent studies have partly addressed the question, but a systematic investigation is lacking.
It has namely been suggested that one may increase the precision of estimated diffusion coefficients by maximizing the motion blur using a double-pulse illumination sequence, i.e., short pulse-like illumination and recording of the sample at the very start and end of each time-lapse~\cite{Berglund2010}.
Another study has investigated the effect of adjusting several experimental parameters in more detail~\cite{Michalet2012}, though without explicitly considering the trade-off between the number of frames recorded (the time-series length) and the signal in each frame.
It suggested that for time-independent illumination, the shutter should be left open during the whole time-lapse to maximize photon economy, and the number of photons recorded in an image should be enough to practically maximize the information content in individual recorded frames.
These results relied on assumptions that neglected subtle but fundamental details of localization of diffusing particles.
The former study~\cite{Berglund2010} neglected that motion blur increases the width of the measured photon distribution at the camera [the point-spread function (PSF)], increasing the localization error.
The latter~\cite{Michalet2012} took this effect into account, but neglected background noise, which is inevitable in experiment and leads to a non-linear dependence of the localization error on motion blur; this effect is especially important when motion blur or background noise is high. 

We here perform a systematic analytical and numerical study of how to choose experimental parameters for tracking of freely diffusing particles in order to maximize the information in recorded time-series.
We consider two different scenarios which cover the  experimental situations usually encountered in single-particle tracking: (i) where the time that a particle can be followed, the {\sl recording time}, $\ttot$, is the limiting factor; (ii) where the photostability of the fluorophore, and thus the total number of signal photons, $\Ptot$, that can be recorded is the limiting factor.
We show for both cases how to optimize experimental parameters.

In order to answer the question of how to optimally choose experimental parameters for tracking of diffusing particles, we first need to study how motion blur and limited photon statistics influence the precision of optimal estimators of the diffusion coefficient.
This is done in Sections~\ref{sec:localization} and \ref{sec:estimation}. Section~\ref{sec:optimal} applies these results to optimize experimental design.
Section~\ref{sec:estimatorChoice} discusses the choice of the estimator of diffusion coefficients in practice.

Specifically, we investigate in Section~\ref{sec:localization} how limited photon statistics and motion blur affect the precision of commonly used localization methods. We review analytical results for the localization error that ensues when  localizing a stationary particle. We then derive an expression for the average measured width of the PSF of a diffusing particle. Using this, we give an approximate expression for the localization error for a diffusing particle, valid when the mean diffusion length of the tracked particle is smaller than the width of the PSF of a stationary fluorophore.

In Section~\ref{sec:estimation}, we next review the statistics of time-lapse recorded data of a freely diffusing particle and use the results of the previous section to investigate how motion blur affects the precision of estimates of diffusion coefficients.
We show that recording using the double pulse illumination sequence suggested in~\cite{Berglund2010} tends to {\sl increase} the error on diffusion coefficient estimates.
However, when recording with time-independent illumination and leaving the shutter open continuously, the effect of motion blur is negligible for relevant values of experimental parameters, and focus should be on photon economy.

Building on these results, we  show in Section~\ref{sec:optimal} how experiments should be optimized for maximum precision in the different experimental scenarios.
In general, experiments should be designed to maximize the number of frames recorded, not the number of photons recorded per frame---only enough photons should be recorded such that localization does not fail.
This maximizes the information content in the time-series and, in turn, the precision of estimated diffusion coefficients.
The reason for this is that the precision of estimates of the diffusion coefficient increases as the square root of the number of recorded positions, while the decrease in the signal in individual frames does not influence the precision as much, except for very low signal where localization will tend to fail.

Section~\ref{sec:estimatorChoice} finally gives a brief discussion of how to estimate diffusion coefficients in practice from optimally recorded trajectories. We show that this is done optimally using the regression-free covariance-based estimator (CVE) of~\cite{Vestergaard2014}.

Details on how the precision of the various localization methods was characterized on Monte Carlo generated data is found in Appendix~\ref{app:MC}, and supplemental figures are found in Appendix~\ref{app:supFigs}.

\section{Localizing a diffusing particle}
\label{sec:localization}
In this section, results for localization in single-particle tracking are reviewed and the influence of motion blur and limited photon statistics is investigated.
We consider in the following only diffusion in the image plane. However, for typical particle tracking experiments, where the
focal plane is kept the same throughout the experiment (i.e., focus is not changed to follow an individual particle), we show that diffusion along the optical axis effectively contributes to the localization error simply by a constant additive term and a slight change of the motion blur coefficient.
This means that conclusions drawn here for 2D diffusion in the image plane also hold for 3D diffusion.

In Subsection~A we review localization of stationary particles and give expressions for the localization error associated with different methods.
In Subsection~B we then derive an expression for the average width of the PSF of a diffusing particle for a general time-dependent shutter/illumination sequence.
Finally, in Subsection~C, following the same approach as~\cite{Deschout2012}, we use this result to extend the expressions for localization error to tracking of a diffusing particle. We compare the analytical results to Monte Carlo simulations and discuss the limits of the analytical approach.

\subsection{Localization error for a fixed particle}
\label{sec:LocStat}
The diffraction-limited PSF emitted by a freely rotating fluorescent molecule or a fluorescent bead recorded by a CMOS, CCD, or EMCCD camera is well approximated by a two-dimensional (2D) Gaussian function plus a constant background term~\cite{Mortensen2010}. For an isolated fluorophore of this kind with fixed position, fitting a 2D Gaussian plus a constant to the measured PSF allows us to estimate the position of the molecule more precisely than the width of the PSF.
This is done optimally using the maximum likelihood estimator with Gaussian PSF (MLEwG)~\cite{Mortensen2010}.

When the fluorescent particle's position is estimated using MLEwG it leads to a white-noise localization error with variance
\begin{equation}
  \s_0^2 = \frac{Fs_a^2}{P}\left(1+\int_0^1 \frac{\ln t}{1+Pa^2t/(2\pi b^2s_a^2)}\,dt \right)^{-1} \enspace,
  \label{eq:PSF_mle}
\end{equation}
for camera pixel width $a$, background photon count $b^2$, total number of {\sl signal photons} in the Gaussian part of the PSF, $P$~\cite{note-signalPhotons}, and effective PSF width $s_a^2 = s_0^2 + a^2/12$, where $s_0$ is the width of the PSF  of a stationary fluorophore (typically $s_0\approx100$--$150\,\mathrm{nm}$~\cite{Mortensen2010,Deschout2012}) and the additive term $a^2/12$ is due to camera pixelation and is independent of microscope magnification~\cite{Mortensen2010}.
Finally, $F$ is a factor describing {\sl excess noise} in the camera: for a CCD or CMOS camera, there is no excess noise and $F=1$;
for and EMCCD camera, the stochastic electron multiplication stage leads to excess noise, i.e., a factor two increase in the variance of photon counts in individual pixels, resulting in $F=2$~\cite{Mortensen2010}.

The particle's position is often estimated by a least squares fit to the PSF---the Gaussian Mask Estimator (GME)~\cite{Mortensen2010}---or by determining the centroid of an area containing the PSF~\cite{Deschout2012}. This results in a localization error with variance of the form~\cite{Mortensen2010,Deschout2012}
\begin{equation}
  \s_0^2 = \frac{Fs_a^2}{P}\left(\alpha + \beta\frac{2\pi b^2s_a^2}{Pa^2}\right)
  \label{eq:PSF0} \enspace,
\end{equation}
where for GME $\alpha=16/9$ and $\beta=4$~\cite{Mortensen2010}, and for the centroid method $\alpha=1$ and $\beta=81/8$ when all pixels contributing to the PSF, and only these, have been included~\cite{Deschout2012}.

In the following we assume a linear relation between the amplitude of the background photon noise and the amplitude of the peak signal, i.e, $b^2/a^2=q\,P/(2\pi s_a^2)$, where $b^2/a^2$ is the density of background photons and $P/(2\pi s_a^2)$ is the density of signal photons at the peak of the PSF. Here $q$ is a proportionality factor, which we shall refer to as the {\sl background-to-signal ratio}.
This accounts for both the background noise from autofluorescence and other fluorophores, as well as the contribution from the power-law tails of the true PSF~\cite{Mortensen2010}. The second can normally be absorbed in the background, but is seen when the background noise is low. The background-to-signal ratio $q$ is typically of the order of one~\cite{Mortensen2010}.
Using this definition of $q$, Eqs.~(\ref{eq:PSF_mle}) and (\ref{eq:PSF0}) can be simplified:
for MLEwG,
\begin{equation}
  \s_0^2= \frac{Fs_a^2}{P} \left(1+\int_0^1 \frac{\ln t}{1+t/q}\,dt \right)^{-1} \enspace,
  \label{eq:MLE0}
\end{equation}
and for GME or the centroid method,
\begin{equation}
  \s_0^2 = \frac{Fs_a^2}{P}\left(\alpha + \beta q\right) \enspace.
  \label{eq:GME-centroid0}
\end{equation}

In practice, when localizing a particle, one must first define a general region of interest (ROI) containing only the particle one wants to track. The choice of the ROI naturally affects the localization precision.
The centroid method is particularly sensitive to this as including pixels that only contain background noise increases its error---the error continues to increase as more background pixels are included, diverging with the size of the ROI.
GME and MLEwG, which fit the background noise as a constant offset, are less sensitive to background noise and thus to the size of the ROI.
However, errors in correctly defining the ROI will adversely affect the performance of any localization method.
Common procedures for defining the ROI involve a thresholding procedure~\cite{Deschout2012,Braeckmans2010}, which only retains pixels with a photon count over a certain threshold and selects the largest cluster of such pixels as the ROI (Appendix~\ref{app:MC}).
Correctly determining the ROI notably becomes difficult when signal photons are few.

\begin{figure}
  \includegraphics{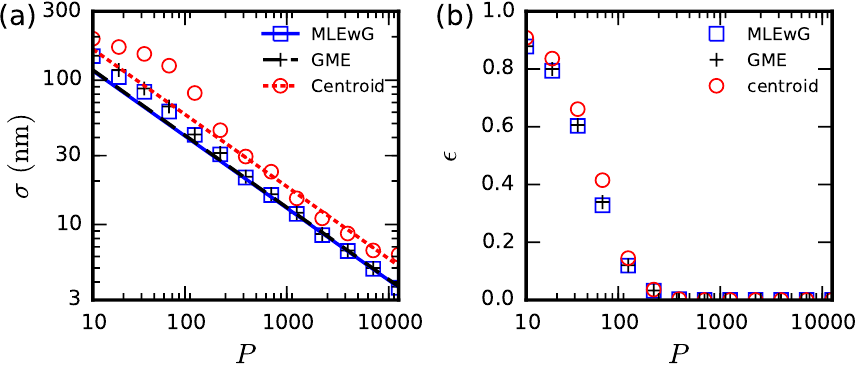}
  \caption{(Color online) Performance of the various localization methods for a static particle as function of the number of signal photons recorded, $P$.
  (a) Amplitude of localization errors of the various methods.
  (b) Probability of failure, i.e., fraction of cases where the localization procedure fails to localize the particle (defined as when the error between estimated and true average positions is larger than $3s_a$)~\cite{note-fail}.
  Lines show theoretical results [Eqs.~(\ref{eq:PSF_mle})--(\ref{eq:GME-centroid0})]; symbols show numerical results (Appendix~\ref{app:MC}); error bars are smaller than symbol sizes.
  To produce the plots the following parameter values were used: 
  $s_a=153\,\nm$ and $q=1$.}
  \label{fig:sig(P)}
\end{figure}

We investigate in Fig.~\ref{fig:sig(P)} how limited photon statistics affects the precision of the various localization methods in practice.
The localization error is approximately proportional to $1/\sqrt{P}$ as predicted theoretically; for low $P$, it is somewhat higher in practice than theoretical results, which can be expected due to difficulties in defining the ROI and since Eqs.~(\ref{eq:PSF_mle})--(\ref{eq:GME-centroid0}) are only strictly valid in the limit of large $P$ [Fig.~\ref{fig:sig(P)}(a)].

We also see that the localization procedures sometimes simply fail to localize the particle [Fig.~\ref{fig:sig(P)}(b)]. 
The probability of failure, $\epsilon$, is zero for large $P$ and increases abruptly for $P<\Pmin\approx100$. Here $\Pmin$ then defines the minimal number of photons needed for reliable localization, which in general depends on the localization procedure used. More advanced methods, notably methods using the preceding and following positions of a tracked particle to localize it~\cite{Chenouard2014}, may decrease $P_{\min}$.
Conversely, excess noise, which is not present in the simulations of Fig.~\ref{fig:sig(P)}, will tend to make ROI determination harder since it increases the variance of the background noise by a factor two (see~\cite[Supplementary Note 1]{Mortensen2010} for a detailed treatise on how the electron multiplication step of an EMCCD camera affects photon statistics of a recorded image).
The overall behavior of $\epsilon$ as function of $P$ does not change, however: it is practically zero for large $P$ and approaches one for small $P$. (See~\cite{Chenouard2014} for a thorough review of single-particle-tracking algorithms and comparison of their performance.)

Note finally that for the relatively high $q$ we consider here---typical of SPT experiments---the performance of the GME and MLEwG methods are practically indistinguishable (see Appendix~\ref{app:supFigs} for different values of $q$). Thus, we will in the following show only results for centroid and MLEwG localization, but note that results for GME are the same as for MLEwG. (For very low $q$ and high $P$, MLEwG is a factor $\sim\sqrt{2}$ times more precise than GME~\cite{Mortensen2010}.)

\subsection{Motion blur increases the width of the measured point-spread function}
Now consider a fluorescent particle diffusing in the image plane.
A fluorescent molecule emits photons with a fixed rate in a Poisson process.
The photons are collected by the camera during a time-lapse $\dt$ to create an image.

The diffusion length during a time-lapse, $\sqrt{2D\dt}$, is in general much smaller than the microscope's field-of-view. This means that eventual aberrations in the microscope do not change the shape of the PSF over the course of a single time-lapse, and we may assume that the dispersion of photons in the microscope is independent of the particle's position during the time-lapse.
Thus, we can for the moment neglect diffraction and finite photon statistics. The effect of these are added later by convoluting the PSF of a stationary particle with the distribution of positions of the diffusing particle during the time-lapse.
Furthermore, since the motion in the $x$- and $y$-directions of a particle diffusing in a homogeneous medium are independent, the motion along the two axes are identical and can be treated separately as one-dimensional (1D) problems. The result derived here is thus valid for both one- and two-dimensional diffusion, and in the following derivation we consider 1D diffusion only.
Finally, since the photon emission process is independent of the particle's position, we do not need to take fluctuations in photon emission into account to derive the average width of the measured PSF.

We can thus split the time-lapse $\dt$ into $M$ points in time, $\tau_0,\tau_1,\ldots,\tau_M$.  At each time-point $\tau_i$, the generic illumination function $I_i$ determines whether the particle's position is recorded. (It can be considered as an indicator function, which is equal to $1$ for time-points when the particle's position is recorded and is equal to zero otherwise.)
We then get a razor-sharp image of the tracked particle's trajectory.
The width $\delta x$ of the distribution of recorded positions around the center of mass of such a trajectory is given by
\begin{equation}
  \dx = \frac{1}{P}\sum_{i=1}^M I_i(x_i-\overline{x})^2 \enspace,
\end{equation}
where $P=\sum_iI_i$ is the total number of photons recorded and $\overline{x}$ is the average position.
Since $P$ is large (typically of the order of 100 or more) the sum is well approximated by an integral, and the expected value of $\dx$ is
\begin{equation}
  \E{\dx} = \int_0^\dt I(t) \E{[x(t)-\overline{x}]^2}dt \enspace,
  \label{eq:sD2,1}
\end{equation}
where $\overline{x} = \int_0^\dt I(t)x(t)dt$, and $I$ is the continuous illumination function, which satisfies $\int_0^\dt I(t)dt=1$.
We insert the expected value $\E{x(t)x(t')} = 2D\min(t,t')+x(0)^2$ into Eq.~(\ref{eq:sD2,1}) and perform partial integration to get
\begin{widetext}
\begin{eqnarray}
  \E{\dx} &=& 2D\left(\int_0^\dt I(t)dt - \int_0^\dt I(t)\int_0^\dt I(t')\min(t,t')dt'dt \right) \nonumber\\
  &=& 2D\left( -\int_0^\dt I(t)\int_0^t I(t')t'dt'dt + \int_0^\dt I(t)S(t)tdt \right) \nonumber\\
  &=& 2D\int_0^\dt S(t)[1-S(t)]dt \nonumber\\
  &=& 2RD\dt \enspace,
\end{eqnarray}
\end{widetext}
where $R$ is the motion blur coefficient, defined by
\begin{equation}
 R = \frac{1}{\dt}\int_0^\dt S(t)[1-S(t)]dt \enspace,
 \label{eq:R}
\end{equation}
and $S(t)=\int_0^t I(t')dt'$~\cite{Berglund2010}.
Important special cases are:
(i) continuous (time-independent) illumination, used in experiments without stroboscopic setup to maximize photon count---here $R=1/6$;  (ii) an instantaneous illumination pulse, which minimizes motion blur---here $R=0$; (iii) the double pulse illumination suggested in~\cite{Berglund2010}, which maximizes the motion blur---here $R=1/4$~\cite{note-blur}.

Since the photon emission process is independent of the particle's position, the average width of the Gaussian part of the measured PSF (the measured distribution minus the constant background) is
\begin{equation}
  s^2 = s_a^2 + 2RD\dt\enspace.
  \label{eq:PSF}
\end{equation}
For a constant illumination function (i.e. time-independent illumination and continuously open shutter) Eq.~(\ref{eq:PSF}) simplifies to the result found in~\cite{Deschout2012}.
This differs from the result found in~\cite{Michalet2010,Schuster2002} since the initial positions of the particle at the start of each time-frame in those studies were implicitly assumed to be known, which is not the case in actual particle-tracking experiments.

When tracking particles that undergo 3D diffusion, e.g., using confocal microscopy, particles diffuse in and out of focus, which tends to increase the width of the measured PSF. Following the approach of Deschout et al.~\cite[Supporting Material]{Deschout2012}, we may use Eq.~(\ref{eq:R}) to extend their result for the average width of the measured PSF emitted by a fluorescent particle undergoing both in-plane and out-of-plane diffusion,
\begin{eqnarray}
  s^2 &=& s_a^2 + 2RD\dt + s_0^2(z_{\rm lim}^2/3+2RD\dt)/z_0^2 \nonumber\\
  &=& s_a^2 + s_0^2z_{\rm lim}^2/(3z_0^2) + 2R(1+s_0^2/z_0^2)D\dt\enspace.
  \label{eq:PSF3D}
\end{eqnarray}
Here $z_0=4\pi ns_0^2/\lambda\approx4s_0$, with numerical aperture of the objective $n$ and photon wavelength $\lambda$, and $z_{\rm lim}$ is the distance from the focal plane where the particle becomes undetectable.
Diffusion along the optical axis thus simply changes the effective stationary PSF width 
and slightly changes the motion blur coefficient (i.e., by approximately $1/16\approx6\%$). Thus, it does not qualitatively alter the results derived below. We thus consider only diffusion in the image plane in the following, but note that conclusions found here also apply to 3D diffusion.

\begin{figure}
  \centering
  \includegraphics{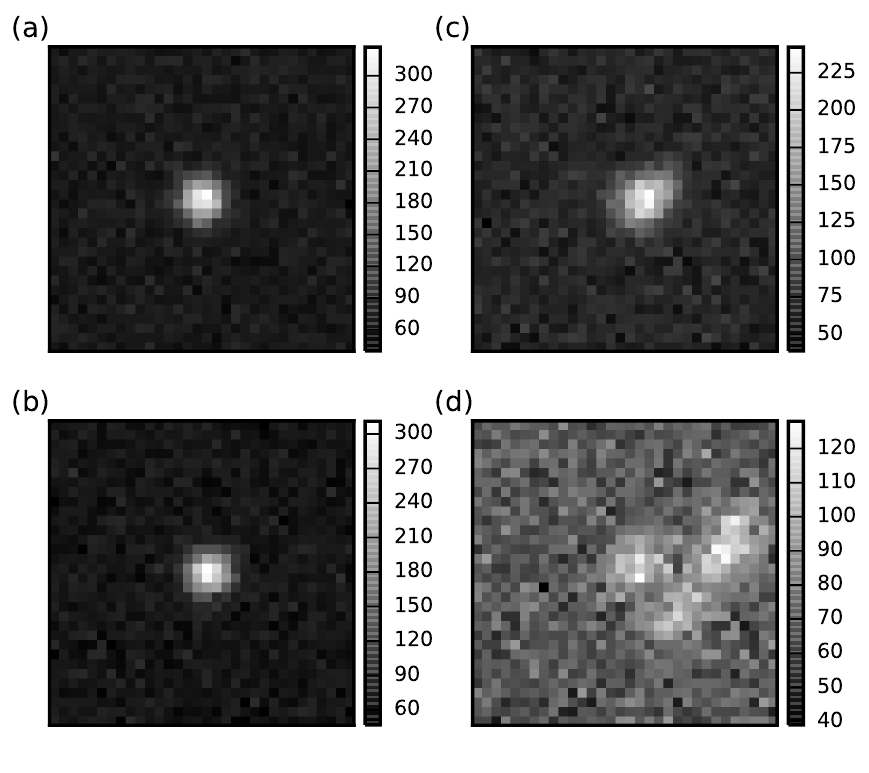}
  \caption{Monte Carlo simulated images of measured PSFs emitted by stationary and diffusing fluorescent point-like particles at low background-noise conditions. (a) Stationary particle. (b)--(d) Particle diffusing in the image plane with mean diffusion length $\sqrt{2D\dt}$ equal to: (b) $\sqrt{2D\dt}=s_a$; (c) $\sqrt{2D\dt}=\sqrt{10}s_a$; (d) $\sqrt{2D\dt}=10s_a$.
  In all images the total number of photons emitted by the particle is $P=10\,000$, the background-to-signal ratio is $q=0.1$, and the pixel size and width of the stationary PSF are $a=100\,\nm$ and $s_0=150\,\nm$, respectively, giving $s_a=153\,\nm$.
  Note that the amplitude of the background noise is the same in the four panels, the difference in scales makes it appear higher in panels (c) and (d).}
  \label{fig:PSF}
\end{figure}

Examples of measured PSFs of a particle diffusing in the image place obtained from Monte Carlo simulations are shown in~Fig.~\ref{fig:PSF}.

\subsection{Localization error for a diffusing particle}
Finally, we use the result derived above to extend expressions for the localization error presented in Sec.~\ref{sec:LocStat} to localization of diffusing particles.

\begin{figure}
  \includegraphics{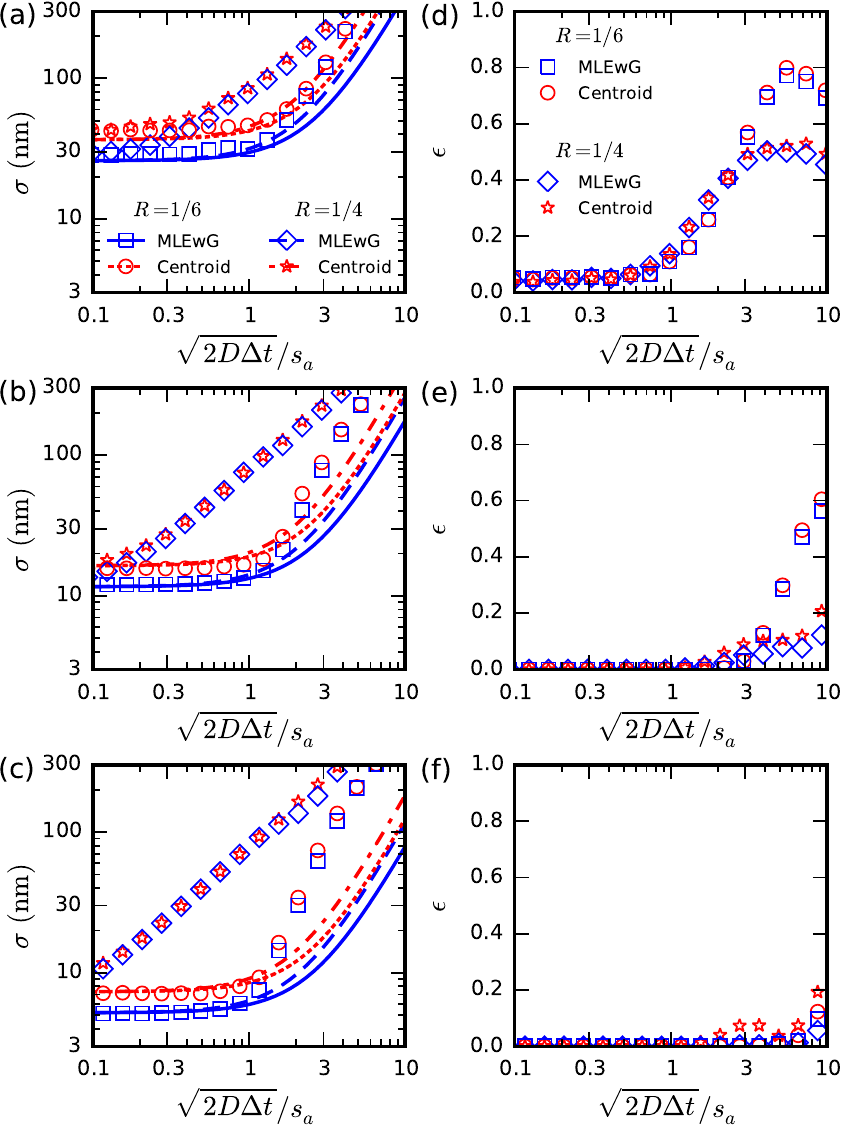}
  \caption{(Color online) Performance of the various localization methods for a diffusing particle as function of 
  $\sqrt{2D\dt}/s_a$.
  (a)--(c) Amplitude of localization errors, $\s$, and (d)--(f) probability $\epsilon$ of localization to fail.
  (The localization is considered to have failed if the error between the estimated and true average positions is higher than $3s=3\sqrt{s_a^2+2RD\dt}$.)~\cite{note-fail}
  Lines show theoretical results [Eqs.~(\ref{eq:MLE}) and (\ref{eq:GME-centroid})]; symbols show numerical results (Appendix~\ref{app:MC}); error bars are smaller than symbol sizes.
  The number of signal photons per image are: (a),(d) 200, (b),(e) 1\,000, and (c),(f) 5\,000.
  The width of the stationary PSF is $s_a=153~\mathrm{nm}$, the background-to-signal ratio is $q=1$, and the results are for 2D diffusion in the image plane.
  Discrepancies between numerical results and theory are due to the asymmetry of the recorded PSF; this effect is particularly strong when recording using the double-pulse illumination sequence ($R=1/4$) since the PSF here quickly becomes highly asymmetric.}
  \label{fig:sig}
\end{figure}

Following the same mean-field approximation used in~\cite{Michalet2012,Deschout2012}, we assume that the effect of motion blur on localization error is found simply by replacing $s_a^2$ by $s^2$ [Eq.~(\ref{eq:PSF})] in Eqs.~(\ref{eq:PSF_mle})--(\ref{eq:GME-centroid0}).
We thus have for MLEwG localization:
\begin{equation}
  \s_0^2= \frac{F(s_a^2+2RD\dt)}{P} \left(1+\int_0^1 \frac{\ln t\,dt}{1+\frac{t}{q(1+{2RD\dt}/{s_a^2})}} \right)^{-1} \enspace,
  \label{eq:MLE}
\end{equation}
while for GME or centroid localization, we have:
\begin{equation}
  \s^2 = \frac{F(s_a^2+2RD\dt)}{P}\left[\alpha + \beta q\left(1+\frac{2RD\dt}{s_a^2}\right) \right] \enspace.
  \label{eq:GME-centroid}
\end{equation}
Equations~(\ref{eq:MLE}) and (\ref{eq:GME-centroid}) show that background noise leads to a faster-than-linear increase in the localization error as function of the motion blur [Fig.~\ref{fig:sig}(a)--(c)].

The above result assumes a symmetrical PSF. However, since diffusion is not a stationary process, the contribution to the measured photon distribution from the diffusive movement is only symmetrical on average, not in a single image.
This means that we can expect the analytical result to break down when motion blur is high enough to make the individual PSF significantly asymmetrical.
In practice, for continuously open shutter ($R=1/6$), the theoretical result agrees well with numerical simulations when the diffusion length during a time-lapse is smaller than $s_a$, i.e., when $\sqrt{2D\dt}/s_a\leq1$ [Fig.~\ref{fig:sig}(a)--(c)].
For experiments using the double-pulse illumination sequence to maximize motion blur, the measured PSF is more asymmetrical, and the localization error is in practice higher than theoretically predicted even when the diffusion length is relatively small.
When $\sqrt{2D\dt}<s_a$ for continuously open shutter ($R=1/6$), the MLEwg and GME estimators are slightly more precise than the centroid estimator.
When recording with double-pulse illumination ($R=1/4$) or for $\sqrt{2D\dt}>s_a$ with continuously open shutter ($R=1/6$), errors in ROI determination dominate the localization error, and all three estimators perform equivalently. Furthermore, increasing $\sqrt{2D\dt}$, especially when photon count is low, increases the probability of the localization procedures to fail [Fig.~\ref{fig:sig}(d)--(f)].

\section{Precision of estimators of the diffusion coefficient from a single trajectory}
\label{sec:estimation}
In this section we build on the results of the previous section to investigate how the precision of estimators of diffusion coefficients depend on experimental parameters. In Subsection~A we review the statistics of time-lapse recorded time-series of diffusing particles and list the parameters that determine the precision of estimators of diffusion coefficients.
In Subsection~B we introduce the Cramér-Rao lower bound (CRB) which limits the precision of any unbiased estimator of the diffusion coefficient---it thus defines the precision of optimal estimators.
In Subsection~C we investigate how the CRB depends on experimental parameters.

\subsection{Statistics of recorded trajectories}
In a single-particle-tracking experiment, a time-series of $N+1$ positions $(\r_0,\r_1,\ldots,\r_N)$ of a particle undergoing isotropic diffusion in $d$ dimensions is determined from images recorded with time-lapse $\dt$.
Each position $\r_n$ is given by
\begin{equation}
  \r_n = \frac{1}{\dt} \int_0^\dt I(t)\r^{({\rm true})}(n\dt-t)dt +\Sigma\bm{\xi}_n \enspace.
  \label{eq:r_n}
\end{equation}
Here $\r^{(\rm true)}$ is the true position of the particle, and the time integral describes motion blur when recording using the illumination function $I$.
The second term describes localization errors associated with the time-averaged position given by the first term, where $\bm{\xi}_n$ is a $d$-dimensional zero-mean Gaussian variable of unit variance and $\Sigma$ describes the amplitude of the localization errors along each coordinate; we define $\s_i=\Sigma_{ii}$.

For diffusion in an isotropic medium, Eq.~(\ref{eq:r_n}) separates into $d$ independent equations describing the motion along each coordinate (up to corrections to the off-diagonal elements of $\Sigma$ due to possible correlations between the amplitude of localization errors along the optical axis and in the image plane).
Thus, the influence of $d$ is trivial---the problem of estimating the diffusion coefficient from the trajectory of a diffusing particle is essentially the same in 1, 2 or 3 dimensions. 
For simplicity, we therefore consider here only $2D$ diffusion in the image plane and assume that $\s_x=\s_y=\s$, but note that conclusions do not depend on this particular choice.

We define the set of $N$ single-time-lapse displacements $(\dr_1,\ldots,\dr_N)$, given by $\dr_n=\r_n-\r_{n-1}$.
These displacements are Gaussian distributed with mean zero,
\begin{equation}
  \E{\dr_n} = 0 \enspace,
  \label{eq:mean(dr)}
\end{equation}
and, for 2D diffusion, with covariance~\cite{Berglund2010}:
\begin{subequations}
  \begin{eqnarray}
    \E{|\dr_n|^2} &=&  4D\dt + 4(\s^2-2RD\dt) \\
    \E{\dr_n\cdot\dr_{n+1}} &=& -2(\s^2-2RD\dt) \\
    \E{\dr_m\cdot\dr_n} &=& 0\ \, \quad \mathrm{for}\ |n-m|>1 \enspace.
  \end{eqnarray}
  \label{eq:cov(dr)}
\end{subequations}
Since free diffusion is translationally invariant, Eqs.~(\ref{eq:mean(dr)}) and (\ref{eq:cov(dr)}) completely characterize the statistics of the recorded trajectory. Thus, the set of single-time-lapse displacements, $(\dr_1,\ldots,\dr_N)$, is a {\sl sufficient statistic} for the trajectory. This means that no estimator, no matter how it uses the information present in the recorded trajectory, can do better than an estimator which optimally uses the information present solely in $(\dr_1,\ldots,\dr_N)$. In particular, it means that the Cramér-Rao bounds derived below limit the precision of {\sl any} unbiased estimator of the diffusion coefficient based on a recorded trajectory.

We define the signal-to-noise ratio $\SNR$ of the trajectory as the mean diffusion length $\sqrt{4D\dt}$ of a particle during one time-lapse divided by the mean contribution $\sqrt{4}\s$ of the localization error to the measured displacement,
\begin{equation}
  \SNR = \frac{\sqrt{D\dt}}{\s} \enspace.
  \label{eq:SNR}
\end{equation}
This signal-to-noise ratio, along with the motion blur coefficient $R$, and the time-series length $N$, determines the precision of any estimator of $D$.
Since $\SNR$ itself depends on $F$, $q$, $P$, the ratio $\sqrt{2D\dt}/s_a$, and $R$, the precision of estimates of $D$ is completely determined by six parameters:
(i) the ratio of the diffusion length to the PSF width, $\sqrt{2D\dt}/s_a$,
(ii) the excess noise factor, $F$,
(iii) the number of signal photons recorded per image, $P$,
(iv) the background-to-signal ratio in images, $q$,
(v) the motion blur coefficient, $R$,
and (vi) the number $N+1$ of frames in the recorded time-series or, equivalently, the time-series length, $N$. 

Excess and background noise, quantified here by $F$ and $q$, influence the precision of particle localization and thus the precision of estimated diffusion coefficients. However, since they do not change with $\dt$, $P$, $R$, and $N$, changing their values will not qualitatively change results, and thus the conclusions presented in this manuscript do not depend on their specific values. We fix in the following $F=1$, corresponding to CDD or CMOS cameras, and $q=1$, corresponding to typical background noise in experiment (other values of $q$ are considered in Appendix~\ref{app:supFigs}).
The parameters we can control in an experiment are typically: $\sqrt{2D\dt}/s_a$ (through $\dt$, where $\sqrt{2D\dt}/s_a\propto\sqrt{\dt}$), $P$ (through both $\dt$ and the photon emission rate, $r$, where $P\propto r\,\dt$), $R$ (by engineering the shutter/illumination sequence), and $N$ (through $\dt$, where $N\propto1/\dt$).

\subsection{Precision of optimal estimators for the diffusion coefficient}
One can construct estimators of the diffusion coefficient $D$ and the variance $\s^2$ of the localization error based on a measured time-series (see e.g.~\cite{Vestergaard2014,Berglund2010,Michalet2012}).
We want such an estimator to be as accurate as possible, preferably unbiased. That is, an estimator of $D$ should on average give the true value of $D$.
Furthermore, we want the estimator to be as precise as possible.
The precision of any unbiased estimator of $D$ is bounded by the information limit, the Cramér-Rao bound (CRB)~\cite{Rao1973}.

An estimator which is unbiased and obtains the CRB is considered optimal---the MLE~\cite{Vestergaard2014,Berglund2010,Michalet2012} does this asymptotically (for $N\to\infty$), and a simple covariance-based estimator (CVE) does this for $\SNR>1$ and all $N$~\cite{Vestergaard2014}.
The commonly used least-squares fitting to measured mean-squared displacements (e.g., as described in~\cite{Sheetz1991}) is suboptimal and its use should be avoided~(for a complete discussion see \cite{Vestergaard2014}).

Technically, the CRB is a lower bound on the variance or, equivalently, the standard error of any unbiased estimator of the set $(D,\s^2)$---or of $D$ alone if $\s^2$ has been determined independently. We here let CRB refer to the lower bound on the standard error, defined as $\Info^{-1/2}$, where $\Info$ is the Fisher information.
When both $D$ and $\s^2$ must be estimated from the time-series, $\Info$ is a matrix given by
\begin{equation}
  \Info = 2\left(\begin{array}{cc}
            \sum_{k=1}^N \frac{1}{\psi_k^{2}}\left(\df{\psi_k}{D}\right)^2  &
           \sum_{k=1}^N \frac{1}{\psi_k^{2}}\df{\psi_k}{D}\df{\psi_k}{\s^2} \\
            \sum_{k=1}^N \frac{1}{\psi_k^{2}}\df{\psi_k}{D}\df{\psi_k}{\s^2} &
            \sum_{k=1}^N \frac{1}{\psi_k^{2}}\left(\df{\psi_k}{\s^2}\right)^2
          \end{array}\right)
          \label{eq:info2}
\end{equation}
with~\cite{Michalet2012}
\begin{equation}
  \psi_k = 2D\dt + 2\left(1-\cos\frac{\pi k}{N+1}\right)(\s^2-2RD\dt) \enspace,
\end{equation}
where $\psi_k$ is the second moment of the normalized discrete sine transform of $(\dr_1,\ldots,\dr_N)$.

It is usually possible to determine $\s^2$ independently, e.g., directly from the localization procedure as described in~\cite{Mortensen2010} when the motion blur is sufficiently small ($\sqrt{2D\dt}< s_a$), or by averaging over estimates of $\s^2$ obtained from multiple time-series recorded under the same experimental conditions, as described in~\cite{Vestergaard2014}.
(Note that the first method relies on an approximately symmetric PSF and thus, if the shutter is held open continuously, that $\sqrt{2D\dt}<s_a$; the second method assumes that the time-series used in the average are recorded with the same localization error, and thus also approximately the same motion blur.)
With $\s$ determined beforehand, all the information in the time-series is used to estimate $D$ alone.
This increases the precision of the estimate~\cite{Vestergaard2014}.
If $\s^2$ has been determined independently with high precision, and this information is used to estimate $D$ from a time-series, the CRB on the standard error of this estimate is
\begin{equation}
  \Info^{-1/2} = D\left[\sum_{k=1}^N \left( \frac{1-2R\left(1-\cos\frac{\pi k}{N+1}\right)}{1+\left(\frac{\s^2}{D\dt}-2R\right)\left(1-\cos\frac{\pi k}{N+1}\right)} \right)^2\right]^{-1/2} \enspace.
  \label{eq:var(D_1)}
\end{equation}

\begin{figure}
  \includegraphics{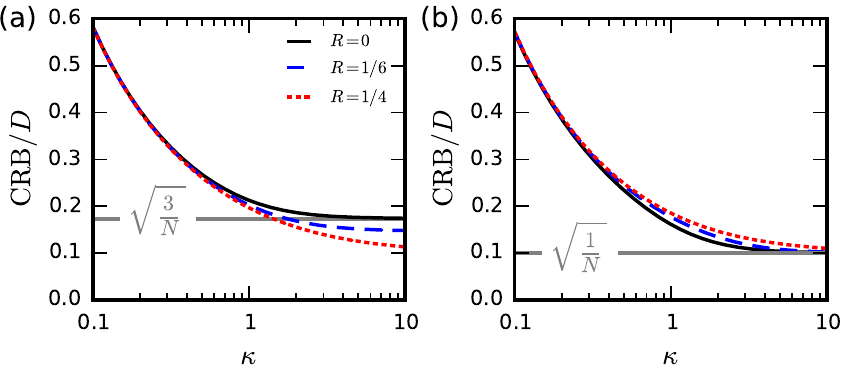}
  \caption{(Color online) Cramér-Rao bound (CRB) on the standard error of any unbiased estimator of the diffusion coefficient $D$ as a function of $\SNR=\sqrt{D\dt}/\s$. Shown for different motion blur: no motion blur ($R=0$), motion blur corresponding to continuously open shutter ($R=1/6$), and maximal motion blur ($R=1/4$). The CRB is measured in units of the true value of $D$. (a) If the amplitude of localization errors, $\s$, is unknown, increasing the motion blur while keeping $\SNR$ constant lowers the standard error of an optimal estimator. (b) For known $\s$, the highest precision is obtained for minimal ($R=0$) motion blur. 
  In both plots, results are for 2D diffusion and the length of the time-series is $N=100$.}
  \label{fig:CR(SNR)}
\end{figure}

Increasing $N$ naturally decreases the CRB ($\CRB\sim1/\sqrt{N}$ for $N\gg1$).
Increasing $\SNR$ also decreases the CRB: for $\SNR\ll1/N\ll1$ the CRB scales with $\SNR$ as $\CRB\sim\SNR^{-2}$~\cite{Michalet2012}, for $1/N<\SNR<1$ we have $\CRB\sim\SNR^{-1/2}$~\cite{Michalet2012}, while for $\SNR\gg1$ the CRB approaches a constant value (Fig.~\ref{fig:CR(SNR)}).
For $R=0$, this asymptotic value for the CRB is equal to $\sqrt{3}\,D/\sqrt{N}$ when both $D$ and $\s^2$ are estimated from the same time-series~\cite{Michalet2012}, and it is equal to $D/\sqrt{N}$ when $\s^2$ has been determined independently~\cite{Vestergaard2014}.
More surprisingly, when both $D$ and $\s^2$ are estimated from a time-series, higher $R$ leads to a lower error~\cite{Berglund2010}---if $\SNR$ is kept the same (for $R=1/4$, the CRB approaches $D/\sqrt{N}$ as $\SNR\to\infty$).
If $\s^2$ has been determined independently, however, the CRB is lowest for minimal motion blur ($R=0$) for all values of $\SNR$.

\subsection{Influence of motion blur}
\begin{figure}[!htb]
  \includegraphics{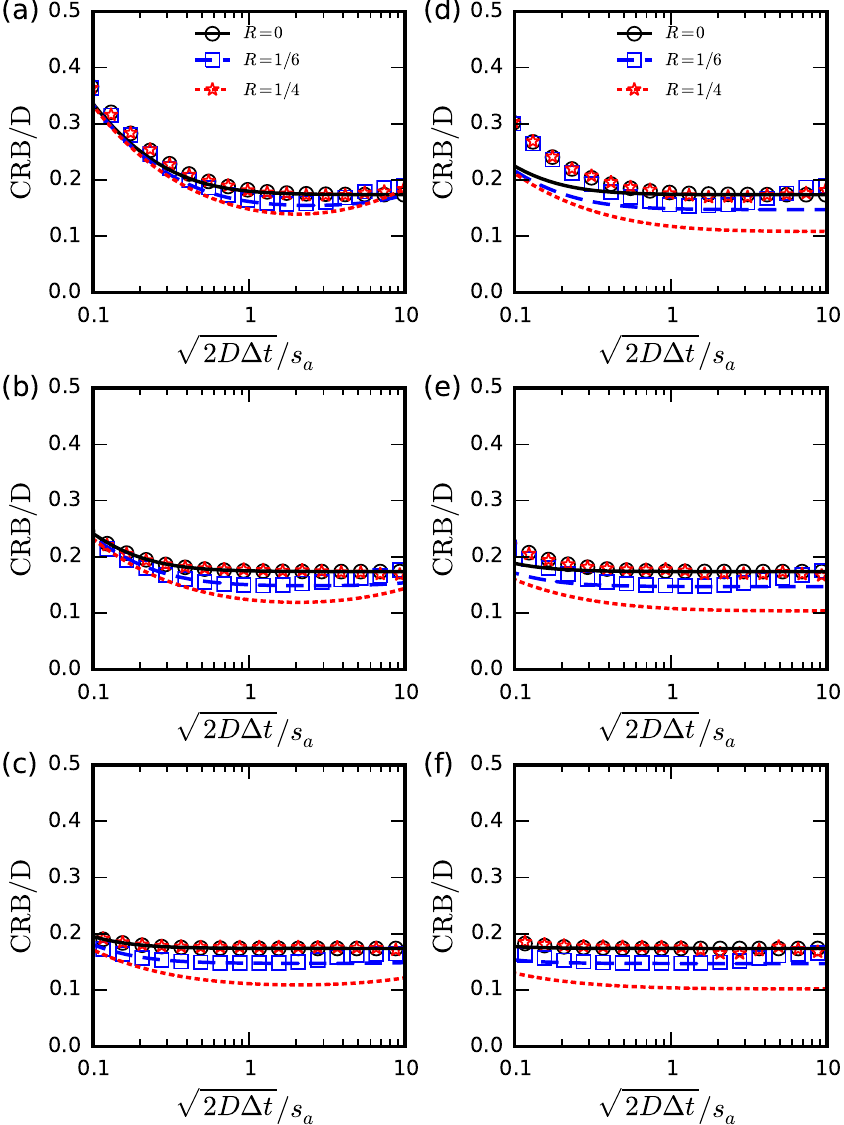}
  \caption{(Color online) Cramér-Rao bound (CRB) on the standard error of any unbiased estimator  of the diffusion coefficient in the presence of motion blur when the amplitude of localization errors, $\s$, is unknown. Results for: no motion blur ($R=0$), continuously open shutter ($R=1/6$), and maximal motion blur ($R=1/4$).
  Lines mark theoretical results [from Eq.~(\ref{eq:info2})]; symbols mark numerical results (Appendix~\ref{app:MC}); error bars are smaller than symbol sizes.
  The particle's positions have been determined using the centroid method (a)--(c) or MLEwG (d)--(f).
  The number of signal photons per image is (a),(d) 200, (b),(e) 1\,000, and (c),(f) 5\,000.
  The motion blur coefficient slightly affects estimator precision, though in general the effect is negligible.
  For low $\sqrt{2D\dt}/s_a$ and $P$, and thus low $\SNR$, the higher localization error of the centroid method leads to a somewhat higher CRB compared to MLEwG.
  In all plots results are shown for 2D diffusion,
  the background-to-signal ratio for photon count is $q=1$, and the time-series length is $N=100$.
  Note that $D$, $\dt$, and $s_a$ do not need to be set to specific values to produce the plots as their ratio $\sqrt{2D\dt}/s_a$ fully determines the CRB.
  The difference between theory and numerical results for $R=1/4$ is due to the high asymmetry of the recorded PSF here.}
  \label{fig:CR}
\end{figure}

\begin{figure}[ht!]
    \includegraphics{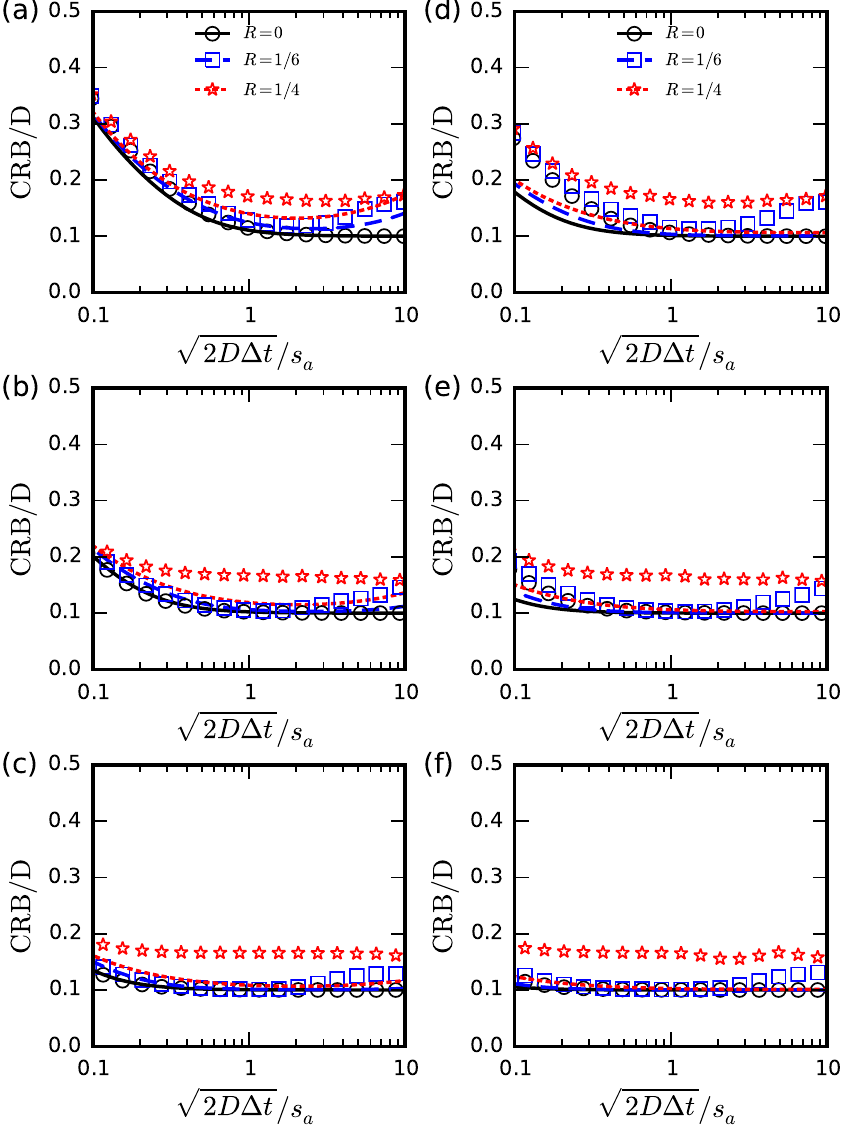}
  \caption{(Color online) Cramér-Rao bound (CRB) on the standard error of any unbiased estimator of the diffusion coefficient in the presence of motion blur when the amplitude of  localization errors, $\s$, has been determined independently.  Results for: no motion blur ($R=0$), continuously open shutter ($R=1/6$), and maximal motion blur ($R=1/4$).
  Lines mark theoretical results [from Eq.~(\ref{eq:var(D_1)})]; symbols mark numerical results (Appendix~\ref{app:MC}); error bars are smaller than symbol sizes.
  The particle's positions have been determined using the centroid method (a)--(c) or MLEwG (d)--(f).
  The number of signal photons per image is (a),(d) 200, (b),(e) 1\,000, and (c),(f) 5\,000. If $\s$ has been determined independently, increasing the motion blur coefficient always increases the estimation error, although only by a negligible amount, unless using the double pulse illumination sequence to maximize motion blur ($R=1/4$).
  In all plots the tracked particle undergoes 2D diffusion,
  the background-to-signal ratio is $q=1$, and the length of the time-series is $N=100$. Note that $D$, $\dt$, and $s_a$ do not need to be set to specific values to produce the plots as their ratio $\sqrt{2D\dt}/s_a$ fully determines the CRB.
  The difference between theory and numerical results for $R=1/4$ is due to the high asymmetry of the recorded PSF here.}
  \label{fig:CR_s2}
\end{figure}

The surprising conclusion that engineering the experiment (through the choice of $I$) to maximize the motion blur coefficient, $R$, may decrease the error on estimated diffusion coefficients hinges on the implicit, yet crucial, assumption that changing $R$ does not change $\SNR$. However, as we have seen in Sec.~\ref{sec:localization}, increasing $R$ increases the localization error and consequently decreases $\SNR$; especially if one uses the double pulse illumination sequence to maximize $R$ (Fig.~\ref{fig:sig}).
So, one should compare estimator precision, not for fixed $\SNR$, but for fixed physical parameters, which are not affected by the choice of $I$: the diffusion coefficient $D$, the PSF width $s_a$, the time-lapse $\dt$, and the time-series length $N$~\cite{note-P(R)}.
Inserting Eqs.~(\ref{eq:MLE}) and (\ref{eq:GME-centroid}) into Eq.~(\ref{eq:SNR}) gives:
\begin{equation}
  \SNR = \sqrt{\frac{f\left(q\left[1+\frac{2RD\dt}{s_a^2}\right]\right)D\dt\,P}{F\,(s_a^2+2RD\dt)}} \enspace,
  \label{eq:SNR_physical}
\end{equation}
where $f(x)=(1+81x/8)^{-1}$ for centroid localization, $f(x)=(16/9+4x)^{-1}$ for GME, and $f(x)=1+\int_0^1 \ln t/(1+t/x)\,dt$ for MLEwG.
[Since Eqs.~(\ref{eq:MLE}) and (\ref{eq:GME-centroid}) are accurate for $\sqrt{2D\dt}/s_a\leq1$, Eq.~(\ref{eq:SNR_physical}) is also accurate for $\sqrt{2D\dt}/s_a\leq1$.]
Taking into account the motion blur's influence on the localization error reveals that, while maximizing $R$ as suggested in~\cite{Berglund2010} may in theory increase precision of estimated diffusion coefficients, in practice it {\sl decreases} the precision due to its detrimental effect on localization precision (Figs.~\ref{fig:CR} and \ref{fig:CR_s2}).

For all $R>0$, the CRB depends non-monotonously on $\sqrt{2D\dt}/s_a$ (Figs.~\ref{fig:CR} and \ref{fig:CR_s2}).
As $\sqrt{2D\dt}/s_a$ is increased, the CRB first decreases as $\SNR$ increases.
For intermediate $\sqrt{2D\dt}/s_a$, the CRB stays constant at its minimal value since further increase in $\SNR$ does not influence the CRB when $\SNR$ is much larger than one.
For high $\sqrt{2D\dt}/s_a$, the CRB increases again; this is due to $\SNR$ eventually decreasing since the localization error increases faster than linearly with the motion blur [Eqs.~(\ref{eq:MLE}) and (\ref{eq:GME-centroid})].
If one also considers that motion blur may induce failure of the localization procedure~[Figs.~\ref{fig:sig}(d)--\ref{fig:sig}(f)], which effectively reduces $N$, the increase in the CRB due to high motion blur is more dramatic~(Appendix~\ref{app:supFigs}).
This is in contrast to the result of~\cite{Michalet2012}, where the omission of background noise led to the prediction that $\SNR$ would asymptotically approach maximum, and thus that the CRB would approach minimum, as $\sqrt{2D\dt}/s_a$ is increased.

For moderate values of $\sqrt{2D\dt}/s_a$, however, the CRB depends little on $R$. In particular, recording with continuously open shutter ($R=1/6$) is optimal in practice as long as $\sqrt{2D\dt}/s_a$ is on the order of one or smaller.
(For a typical value of the diffusion coefficient of $D\approx1\,\mum^2\sec^{-1}$, this means that $\dt$ should be smaller than $s_a^2/(2D)\approx10\,{\rm ms}$; for different values of $D$, this bound changes as $\propto1/D$, e.g., for $D=0.1\,\mum^2\sec^{-1}$: $\dt<100\,{\rm ms}$, or for $D=10\,\mum^2\sec^{-1}$: $\dt<1\,{\rm ms}$.)

We have above assumed that $P$ is not affected by engineering camera shutter and sample illumination in order to either increase or decrease motion blur (i.e. changing $R$ from the value $R=1/6$).
In general, even if stroboscopic techniques are employed, this typically decreases the number of photons recorded during a frame, i.e., since the shutter is kept closed during part of the time-lapse.
Thus the precision obtained in practice when recording with $R\approx0$ or $R\approx1/4$ can be expected to be lower than shown in Figs.~\ref{fig:CR} and \ref{fig:CR_s2}, tipping the scale further in favor of simply recording with the shutter continuously open.

In summary: as long as $\sqrt{2D\dt}/s_a\lesssim1$, which is the relevant parameter range for optimizing the experiment (we shall see below), leaving the shutter open continuously gives the highest precision of estimated diffusion coefficients.

\section{Optimizing experimental design}
\label{sec:optimal}
Using the results derived in the preceding sections, we show how one should adjust the rate at which the tracked fluorescent particle emits photons, $r$, the time-lapse of recordings, $\dt$, and the motion blur coefficient, $R$, in order to maximize the information about the diffusion coefficient contained in a recorded time-series.
We consider two different experimental scenarios, which together cover the experimental situations usually encountered in single-particle tracking.
In Subsection~A we consider the situation where the time that a particle can be followed, $\ttot$, is limited, e.g., since it diffuses out of the field of view of the microscope.
In Subsection~B we consider the situation where the photostability of the fluorescent particle limits the total number of photons that can be collected for a single trajectory, $\Ptot$.

\subsection{Limited experimental recording time}
The total time a particle can be recorded, $\ttot = (N+1)\dt$, may be limited by factors beyond experimental control.
The particle may for example detach from the substratum on which it diffuses
(e.g. a cellular structure~\cite{Blainey2006}).
Or the particle may diffuse out of the microscope's field-of-view. The
latter is typical for particles diffusing
in three dimensions.
Alternatively, we may suspect the diffusion coefficient to change over time, and we may thus be constrained to determine it from trajectories measured over a short time to test this hypothesis.
If the relevant time-scale is shorter than the time-scale of bleaching, it is clear that one should record with the shutter continuously open ($\SNR$ scales as $\sim\sqrt{P}$, but depends little on $R$ for $R\leq1/6$).
Furthermore, we should maximize photon count by maximizing the rate of photon emission from the fluorescent particle.
The fluorophore's emission rate is limited in practice by the photochemistry of the fluorescent particle, by the laser power that may be used without photodamaging the sample, or due to detector saturation in the camera.

In the following, we thus set $R=1/6$ and set $r$ as high as possible. The performance of estimators of the diffusion coefficient is then determined by the parameters $D$, $s_a$, $\ttot$, and $\dt$.
The parameters $D$ and $\ttot$ are beyond our experimental control,
while $s_a$ is determined by the emission wavelength of the fluorophore, the numerical aperture of the microscope, and the camera resolution~\cite{Mortensen2010}, and in general varies little for typical choices of the three.
This leaves us with choosing $\dt$; we investigate below how this should be done to optimize precision.

\begin{figure}[t!]
  \centering
  \includegraphics{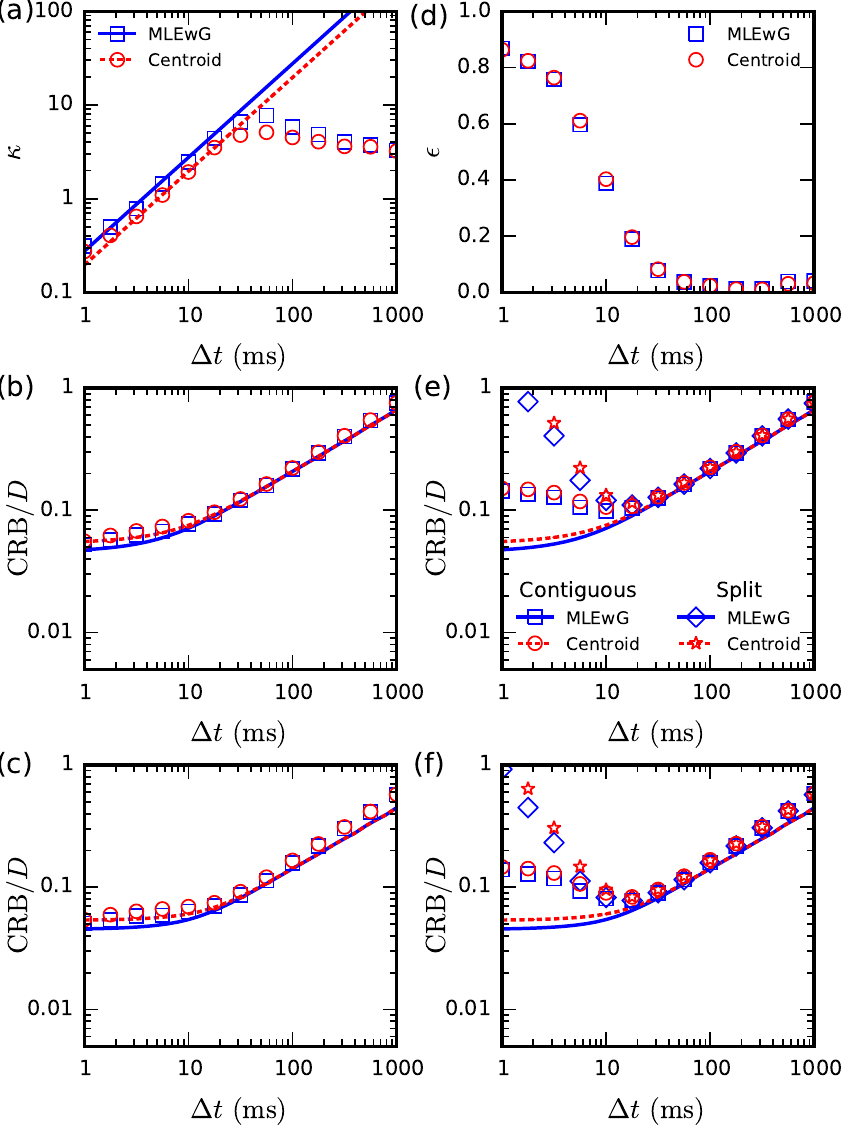}
  \caption{(Color online) Cramér-Rao bound (CRB) on the standard error of any unbiased estimator of the diffusion coefficient as function of time-lapse $\dt$ for a time-series of limited recording time $\ttot=(N+1)\dt$.
  Lines mark theoretical results [from Eqs.~(\ref{eq:info2}) and (\ref{eq:var(D_1)})]; symbols mark numerical results (Appendix~\ref{app:MC}); error bars are smaller than symbol sizes.
  (a) Signal-to-noise ratio, $\SNR$, in a single frame as function of $\dt$.
  (b),(c) CRB as function of $\dt$ when all $N+1$ positions are found: (b) for unknown $\s$, (c) for independently determined $\s$.
  (d) Probability $\epsilon$ for localization to fail as function of $\dt$.
  (e),(f) CRB as function of $\dt$ when the particle is only localized in a fraction $1-\epsilon$ of the $N+1$ images recorded, for the two different scenarios discussed in the main text: (i) where a contiguous trajectory can be constructed from the found positions (Contiguous), or (ii) where the particle cannot be reidentified after a position is missing and the trajectory is split into smaller time-series (Split);
  (e) for unknown $\s$, (f) for known $\s$.
  In all panels the total recording time is $\ttot=10\,\sec$, the rate of photon emission of the fluorescent particle is $r=10\,\mathrm{kHz}$, the width of the stationary PSF is $s_a=153\,\mathrm{nm}$, the background-to-signal ratio in images is $q=1$, the shutter is held continuously open ($R=1/6$), and the particle undergoes 2D diffusion with diffusion coefficient $D=1\,\mum^2s^{-1}$.}
  \label{fig:t_tot}
\end{figure}

For a particle that can be tracked for a time $\ttot$, the number of recorded signal photons per image is $P=r\dt$ and the length of the recorded time-series is $N=\ttot/\dt-1$, assuming that one is able to determine the particle's position in all recorded images. Figures~\ref{fig:t_tot}(b) and \ref{fig:t_tot}(c) show that the information content in a time-series is then maximized by choosing $\dt$ as small as possible, even though this leads to smaller $\SNR$ for individual frames [Fig.~\ref{fig:t_tot}(a)]. The reason for this is that as long as $\SNR>1$, changing it does not change the CRB much (Fig.~\ref{fig:CR(SNR)}), while the CRB always decreases with $N$ as $\CRB\sim 1/\sqrt{N}$~\cite{Michalet2012}.
Specifically, in the relavant range, where $1/N\ll\SNR\ll1$, we have  $\CRB\sim N^{-1/2}\SNR^{-1/2}$ (Fig.~\ref{fig:CR(SNR)}), and since $\SNR\sim\dt$~[Eq.~(\ref{eq:SNR_physical}) and Fig.~\ref{fig:t_tot}(a)] and $N\propto\dt^{-1}$, the CRB asymptotically approaches minimum as $\dt$ is decreased towards zero. 
Note that the results of Figs.~\ref{fig:t_tot}(b) and \ref{fig:t_tot}(c) hinge entirely on the scaling discussed above. They do thus hold, regardless of the values of physical parameters.

In practice, since the number of signal photons recorded per time-lapse, $P$, decreases with $\dt$, the probability for the localization procedure to fail, $\epsilon$, increases as $\dt$ is decreased~[Fig.~\ref{fig:t_tot}(d)].
(Better performing tracking methods, which e.g. use information on the particle's position inferred from preceding and following images, may need fewer photons per image for correct localization; nonetheless, all localization methods, no matter their performance, eventually fail as $P$ decreases.)
Thus, the number of recorded positions is not $(N+1)$ when $\dt$ is small, but rather $(1-\epsilon)(N+1)$. This means that one cannot hope to obtain the ideal results of Figs.~\ref{fig:t_tot}(b) and \ref{fig:t_tot}(c) in practice.

If the density of fluorescent particles is not too high, one is usually able to identify the full trajectory of a particle, even if some positions are missing. One may then fit the resulting time-series with an estimator that accounts for missing positions (see Sec.~\ref{sec:estimatorChoice})~\cite{Shuang2013,Relich2015,Vestergaard2016}.
In this case, the time-lags between frames are not all equal to $\dt$ but are instead equal to integer multiples of $\dt$. This means that Eqs.~(\ref{eq:info2}) and (\ref{eq:var(D_1)}) are no longer valid here. The results of Figs.~\ref{fig:t_tot}(b) and \ref{fig:t_tot}(c) do, however, provide lower bounds on the actual CRB; optimal estimators from the intermittent trajectory are thus at most this precise. 
Conversely, since the signal (the diffusion length) is higher for longer time-lags, the information contained in time-series with missing positions is higher than the information contained in a time-series of the same length (same number of positions) with all time-lags equal to $\dt$. Thus, upper bounds on the CRB can be found from Eqs.~(\ref{eq:info2}) and (\ref{eq:var(D_1)}) with $N$ replaced by $(1-\epsilon)N$; optimal estimators from the intermittent trajectory are at least this precise [``Contiguous'' in Figs.~\ref{fig:t_tot}(e) and \ref{fig:t_tot}(f)].

If one is unable to identify the tracked particle again after a position is missing, one is then left with multiple shorter time-series. In this case, the CRB is given by $(\sum_{m=1}^M\Info_m)^{-1/2}$ [``Split'' in Figs.~\ref{fig:t_tot}(e) and \ref{fig:t_tot}(f)], where $\Info_m$ is the Fisher information for the $m$th individual contiguous time-series, given by Eqs.~(\ref{eq:info2}) or (\ref{eq:var(D_1)}) with $N$ replaced by the length $N_m$ of the individual time-series, and $M$ is the number of such time-series.

In both cases, the CRB initially decreases as $\dt$ is decreased, following the theoretical prediction.
When $P$ becomes too small, the localization procedure starts to fail, and the CRB rapidly increases when $\dt$ is decreased further since a smaller and smaller fraction of positions of the particle are found. The sweet spot, where the CRB is minimal---and precision thus is maximal---is found right before localization fails for a substantial fraction of the recorded images. Here the optimal choice of $\dt$ is around $\dt=10~\ms$, corresponding to $P\approx\Pmin\approx100$.

The position of the optimum depends on the scaling relations discussed above and on the onset of localization failure. The latter depends strongly on $P$, but is insensible to other experimental parameters (for $\sqrt{2D\dt}/s_a$).
Thus, the optimal precision is always found when $P\approx\Pmin$ (Appendix~\ref{app:supFigs}).

This result differs from the recommendation of~\cite{Michalet2012}, which did not consider explicitly the tradeoff between $\dt$ and $N$, and suggested choosing $\dt$ in order to record enough photons in each single image to ensure that $\SNR>1$, and then maximizing the number of frames recorded, $N$, under this constraint.


\subsection{Limited photostability}
Experiments may be limited by the photostability of the fluorescent particle, typical for proteins tagged with, e.g., GFP or an organic dye and bound in a lipid membrane~\cite{Domanov2011}. In this case photon economy is paramount.
Let $\Ptot$ be the total number of photons emitted by the fluorescent particle before bleaching. Then $N=\Ptot/P-1$, with $P$ the number of photons recorded per image.
In principle, an optimal strategy would be to use a stroboscopic setup to record the particle's position during each frame using a short pulse of length $\tau$, where $\tau$ is chosen such that the number of recorded photons $P=r\tau$ is equal to $\Pmin$, and let the time-lapse between images be very long in order to have $\SNR\gg1$.

In practice, less is needed, however.
From Fig.~\ref{fig:CR(SNR)} we know that $\SNR$ needs only be slightly higher than one (or two if $\s^2$ is determined independently) for estimates of the diffusion coefficient from a given time-series to be maximally precise.
We may thus simply choose $\dt$ large enough such that $\SNR$ is always larger than one, i.e., $\dt>\s^2/(2D)$, but small enough to avoid the deleterious effects of high motion blur.
Choosing $\dt\approx s_a^2/(2D)$ accomplishes both in practice. It limits the negative effect of motion blur (Figs.~\ref{fig:CR} and \ref{fig:CR_s2}) and, since we in practice always have $\s<s_a$, it assures that $\SNR>1$ (for $\dt=s_a^2/(2D)$, we have from Eq.~(\ref{eq:SNR_physical}) that $\SNR=\sqrt{P\,f(q[1+R])/[2F(1+R)]}$; so for $P>\Pmin\approx100$, we have $\SNR\gtrsim3$ when localizing using MLEwG or GME and $\SNR\gtrsim2$ when using the centroid method).
For typical values of physical parameters, $D=1\,\mum^2\sec^{-1}$ and $s_a=150\,\nm$, we should thus choose $\dt\approx10\,{\rm ms}$, i.e., a video-rate of 100\,Hz 
(
similarly, for  $D=0.1\,\mum^2\sec^{-1}$ one should choose $\dt\approx100\,{\rm ms}$, while for $D=10\,\mum^2\sec^{-1}$ one should choose $\dt\approx1\,{\rm ms}$, if possible).

When recording with continuously open shutter, $\tau=\dt$ and the performance of estimators of the diffusion coefficient is thus determined by the parameters $\Ptot$, $D$, $s_a$, $\dt$ and $r$.
The parameters $\Ptot$ and $D$ are beyond our experimental control, while $s_a$ is determined by the fluorophore, microscope and camera, and $\dt$ is fixed by our choice to set $\dt=s_a^2/(2D)$.
This leaves us with choosing $r$, which may be controlled experimentally by adjusting the power of the illuminating laser.

\begin{figure}[t!]
  \centering
  \includegraphics{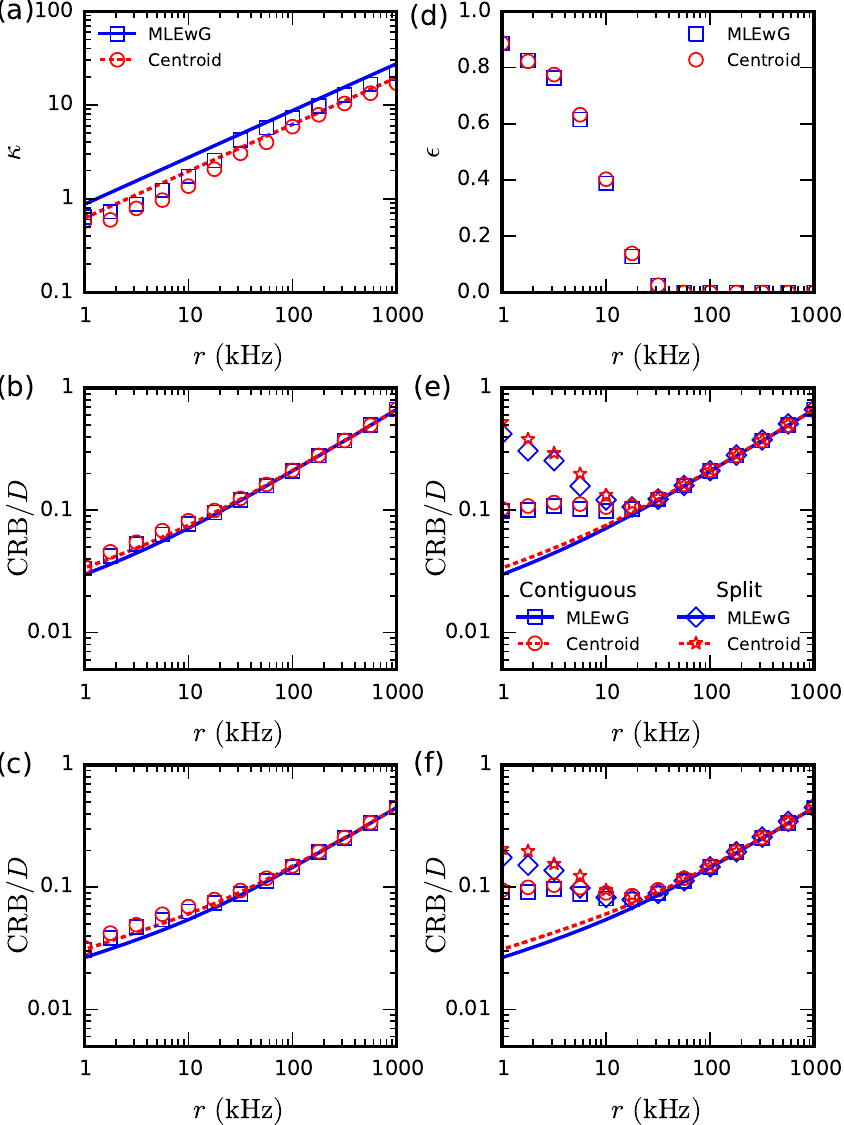}
  \caption{(Color online) Cramér-Rao bound (CRB) on the standard error of any unbiased estimator of the diffusion coefficient as function of photon emission rate $r$ for a time-series whose length is limited by the total number of photons that can be recorded before the fluorophore bleaches, $\Ptot=(N+1)P=(N+1)r\dt$.
  Lines mark theoretical results [from Eqs.~(\ref{eq:info2}) and (\ref{eq:var(D_1)})]; symbols mark numerical results (Appendix~\ref{app:MC}); error bars are smaller than symbol sizes.
  (a) Signal-to-noise ratio, $\SNR$, in a single frame as function of $r$.
  (b),(c) CRB as function of $r$ when the particle is localized in all $N+1$ recorded images:
  (b) for unknown $\s$, (c) for known $\s$.
  (d) Probability $\epsilon$ for localization to fail as function of $r$.
  (e),(f) CRB as function of $r$ when the particle is only localized for a fraction $1-\epsilon$ of the $N+1$ images recorded, for the two different scenarios discussed in the main text: (i) where a contiguous trajectory can be constructed from the found positions (Contiguous), or (ii) where the particle cannot be reidentified and the trajectory is split into smaller time-series (Split):
  (e) for unknown $\s$, (f) for known $\s$.
  In all panels the total number of recorded photons is $\Ptot=10^5$, the time-lapse of recordings is $\dt=10~\mathrm{ms}$, the width of the stationary PSF is $s_a=153~\mathrm{nm}$, the background-to-signal ratio in images is $q=1$, the shutter is held continuously open ($R=1/6$), and the particle undergoes 2D diffusion with diffusion coefficient $D=1~\mum^2s^{-1}$.}
  \label{fig:P_tot}
\end{figure}

In general, as above, maximizing the number of images recorded is more important than maximizing the information in each image, so smaller $r$ lead to higher precision.
Since changing $r$ does not change $2D\dt/s_a^2$, $\SNR$ scales with $r$ as $\SNR\sim \sqrt{r}$ [Fig.~\ref{fig:P_tot}(a)], and since the CRB scales as $\CRB\sim N^{-1/2}\SNR^{-1/2}$  for $1/N\ll\SNR\ll1$, while $N\sim r^{-1}$, we thus have $\CRB\sim r^{1/4}$ [Figs.~\ref{fig:P_tot}(b) and \ref{fig:P_tot}(c)]. That is, the CRB asymptotically approaches zero as $r\to0$. So, in theory, we should let $r$ tend to zero in order to optimize our experiment.

In practice, as $r$ is decreased, $P$ decreases too, which eventually leads to failure of the localization procedure [Fig.~\ref{fig:P_tot}(d)]. As for the case of limited recording time, the probability for localization to fail increases abruptly for $P<\Pmin\approx100$.
If one is able to determine the full trajectory of a particle in spite of the missing positions, Eqs.~(\ref{eq:info2}) and (\ref{eq:var(D_1)}) give lower bounds on the CRB; estimators of the diffusion coefficient from the intermittent trajectory are at most this precise. Conversely, Eqs.~(\ref{eq:info2}) and (\ref{eq:var(D_1)}) with $N$ replaced by $(1-\epsilon)N$ give upper bounds on the CRB; optimal estimators from the intermittent trajectory are at least this precise [``Contiguous'' in Figs.~\ref{fig:P_tot}(e) and \ref{fig:P_tot}(f)].
If one is unable to identify the tracked particle again after a position is missing, the CRB is given by $(\sum_{m=1}^M\Info_m)^{-1/2}$ [``Split'' in Figs.~\ref{fig:t_tot}(e) and \ref{fig:t_tot}(f)], where $\Info_m$ is the Fisher information for the $m$th individual contiguous time-series, given by Eqs.~(\ref{eq:info2}) and (\ref{eq:var(D_1)}) with $N$ replaced by the length $N_m$ of the individual time-series, and $M$ is the number of such time-series.

In both cases, the photon emission rate, $r$, should thus be adjusted to be as low as possible without the localization procedure failing. This means that one should choose $r$ such that the number of recorded signal photons per image is $P\approx\Pmin$ in order to obtain optimal precision of estimates of the diffusion coefficients. For the parameter values of Fig.~\ref{fig:P_tot}, $r\approx10~{\rm kHz}$ is optimal, corresponding to $\Pmin\approx100$.
As in the case of limited experimental recording time, the onset of localization failure determines the position of the optimum. Since this depends strongly on $P$ but is largely insensible to other experimental parameters, optimal precision is obtained when $P\approx\Pmin$ regardless of the values of the physical parameters (Appendix~\ref{app:supFigs}).

\section{Estimating the diffusion coefficient in practice}
\label{sec:estimatorChoice}
The results of the previous section tell us how to choose experimental parameters in order to maximize the information in recorded time-series. This guarantees maximum precision of estimated diffusion coefficients when using an estimator which attains the CRB (i.e. an optimal estimator).
In practice, we need to make a concrete choice of estimation method.
As discussed in detail in~\cite{Vestergaard2014}, common methods based on measured mean-squared displacements (MSDs) squander information and can be expected to perform less than optimally.
Maximum-likelihood estimators (MLEs) are guaranteed to approach optimality in the long time-series limit ($N\to\infty$). They are not guaranteed to be unbiased, however, and for short time-series the bias may be substantial (it scales as $N^{-1/2}$)~\cite{Vestergaard2014,Berglund2010}.
The covariance-based estimator proposed in~\cite{Vestergaard2014} is unbiased by construction and is practically optimal for $\kappa>1$. The CVE is furthermore regression-free, which makes its implementation orders of magnitude faster than MLE and MSD-based methods and allows us to calculate its standard error exactly.

The CVE of \cite{Vestergaard2014} may be adapted in a straightforward manner to estimate the diffusion coefficient from a trajectory with missing positions, we show here.
We define a new index $m=0,1,\ldots,M$ that enumerates the frames in which the particle is successfully localized, with $M\approx N\epsilon$, and we let $\dt_m$ denote the time-lag between frame $m-1$ and $m$. 

For diffusion in 2D, and if the value of $\s^2$ is not known {\sl a priori} so both $D$ and $\s^2$ must be estimated from the time-series, the CVE defined as
\begin{equation}
  \hat{D} = \frac{\overline{|\dr_m|^2}}{4\overline{\dt_m}} + \frac{\overline{\dr_{m+1}\cdot\dr_m}}{2\overline{\dt_m}} \enspace,
  \label{eq:Dcve2}
\end{equation}
where $\overline{\cdots}$ denotes the average of $\cdots$, is an unbiased estimator of $D$.
When the CVE given by Eq.~(\ref{eq:Dcve2}) is used to estimate $D$ , it leads to a standard error of its estimate of $D$ given by~\cite{Vestergaard2016}:
\begin{equation}
  \SE{\hat{D}} = D\sqrt{ \frac{3\overline{l_m^2}+2\overline{l_m}\varepsilon+\varepsilon^2} {N\left(\overline{l_m}\right)^2} + \frac{2\overline{\left(l_m+\varepsilon\right)^2}} {N^2\left(\overline{l_m}\right)^2} } \enspace,
  \label{eq:varCVE2}
\end{equation}
where $\varepsilon=(\s^2-2RD\dt)/(D\dt)=\SNR^{-2}-2R$, and $l_m=\dt_m/\dt$. Here $l_m=1$ for all $m$ if the particle is successfully localized in all recorded frames.

If $\s^2$ has been estimated independently beforehand or is known {\sl a priori}, $D$ may be estimated from an intermittent time-series using
\begin{equation}
  \hat{D} = \frac{\overline{|\dr_m|^2}-4\s^2}{4(\overline{\dt_m}-2R\dt)} \enspace.
  \label{eq:Dcve1}
\end{equation}
In this case the standard error of the CVE's estimate of $D$ is given by~\cite{Vestergaard2016}:
\begin{equation}
  \SE{\hat{D}} =   D\sqrt{\frac{\overline{l_m^2}+2\overline{l_m}\varepsilon+3\varepsilon^2/2} {N\left(\overline{l_m}-2R\right)^2}} \enspace,
  \label{eq:varCVE1}
\end{equation}
where we have assumed that the error on the estimate of $\s^2$ is negligible.

\begin{figure}
  \includegraphics{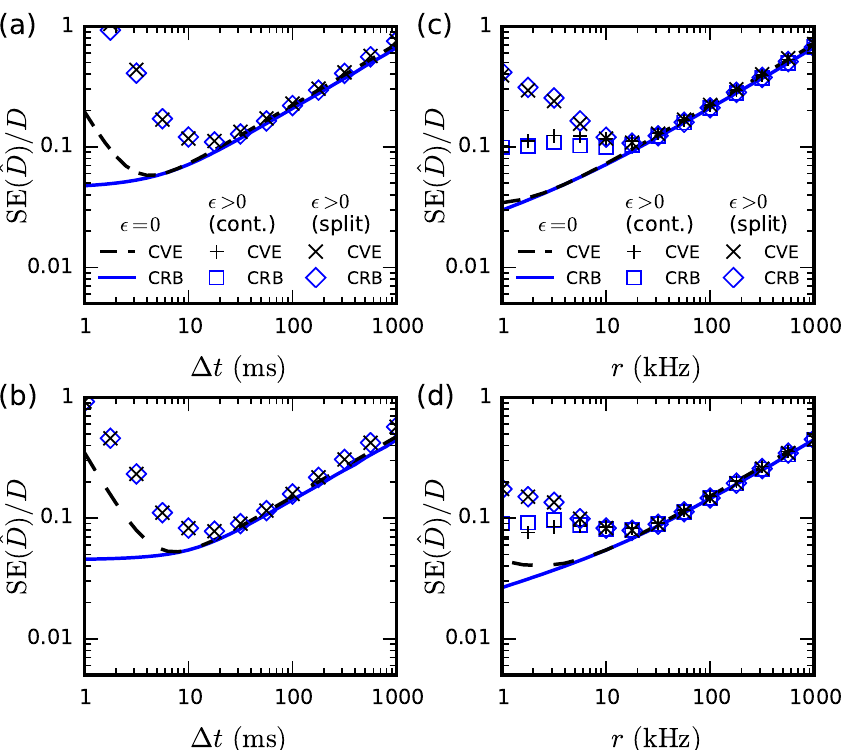}
  \caption{(Color online) Standard error (SE) of the covariance-based estimator (CVE) of the diffusion coefficient~\cite{Vestergaard2014} compared to the Cramér-Rao bound (CRB) for the cases where the tracked particle is successfully localized in all frames ($\epsilon=0$), and where localization of the particle fails for a fraction $\epsilon>0$ of the frames resulting in either a contiguous trajectory with missing positions (cont.) or in multiple shorter trajectories (split).
  Lines mark theoretical results [Eqs.~(\ref{eq:info2}) and (\ref{eq:var(D_1)}) for CRB, Eqs.~(\ref{eq:varCVE2}) and (\ref{eq:varCVE1}) for CVE]; symbols mark numerical results (Appendix~\ref{app:MC}); error bars are smaller than symbol sizes.
  (a),(b) Experimental scenario of limited recording time.
  (c),(d) Experimental scenario of limited fluorophore photostability.
  (a),(c) For unknown localization error variance, $\s^2$; (b),(d) for independently determined $\s^2$.
  In all panels the width of the stationary PSF is $s_a=153~\mathrm{nm}$, the background-to-signal ratio in images is $q=1$, the shutter is held continuously open ($R=1/6$), and the particle undergoes 2D diffusion with diffusion coefficient $D=1~\mum^2s^{-1}$. In (a),(b) the total recording time is $\ttot=10~\sec$ and the rate of photon emission of the fluorescent particle is $r=10~\mathrm{kHz}$.
  In (c),(d) the total number of recorded photons is $\Ptot=10^5$ and the time-lapse of recordings is $\dt=10~\mathrm{ms}$.}
  \label{fig:CVE}
\end{figure}

We show in Fig.~\ref{fig:CVE} how the precision of the CVE compares to the CRB in practice.
Although the precision of the CVE deteriorates quickly for $\SNR<1$, the error is here dominated by the failure of localization, and for a (near) optimal choice of experimental parameters, the CVE reaches the CRB in practice.

\section{Conclusion}
We have shown that one should choose quantity over quality when it comes to tracking diffusing particles. In general, experiments should be designed with focus on maximizing the number of frames recorded---the time-series length---even if this means a low signal-to-noise ratio for individual frames.
In particular, if the time a particle can be recorded is limited, e.g., by the particle diffusing out of the field-of-view, one should record the particle with a photon emission rate and a video rate that are as high as possible.
If the experiment is limited by the fluorescent particle's photostability, one should minimize the photon emission rate and record with a video rate that is slow enough to maximize the information content in each recorded frame yet fast enough to avoid the deleterious effects of motion blur---this is achieved by choosing the video rate such that the mean diffusion length per time lapse is approximately equal to the PSF width of a stationary particle.
In both cases, the fundamental limit on the precision is set by the minimal number, $\Pmin$, of signal photons needed in a single image for reliable localization.

The exact values of optimal $\dt$ and $r$ depend on experimental and physical parameters of the system under study.
However, the results presented in this paper may be used in one of two following ways in practice.
(i) The quick and dirty way: according to whether recording time or photostability limits time-series length, fix either $r$ or $\dt$ and adjust the other to experimentally determine $\Pmin$ as the point where probability for the localization to fail a becomes significant (e.g. $\epsilon\approx0.01$). The parameters giving this $\Pmin$ are then approximately the optimal choice.
(ii) The thorough way: if one wants to squeeze out every last drop of information from the experiment, one may follow the procedure described in the present paper to numerically find optimal experimental parameters for a given setup and localization method. This may even be done iteratively as $D$ is estimated from experiments.
In practice, diffusion coefficients can be estimated optimally from recorded time-series using the covariance-based estimator (CVE) introduced in~\cite{Vestergaard2014}.

Similar procedures to the one presented here may be used to study how to optimize experimental parameters for tracking particles undergoing more complicated forms of motion, such as persistent random motion, active transport, or anomalous diffusion.
Note that optimization in the definitive sense requires that an optimal estimator exists for the motion studied.
While this often is not the case, at least not yet, one may still optimize experiments for a given (suboptimal) estimator of the motility parameters of the motion under study.

\section*{Acknowledgements}
The author thanks R. Mastrandrea, J. Fournet, and A. Barrat for helpful discussions and H. Flyvbjerg, J. N. Pedersen, K. I. Mortensen, and M. Génois for valuable suggestions and critical reading of the manuscript.
This work was partly supported by the Human Frontier Science Program
Research Grant GP0054/2009-C.

\appendix
\section{Numerical simulations}
\label{app:MC}
The analytical results derived in the present paper rely on two simplifying assumptions.
First, they neglect that in real-world tracking experiments, one  must define a region of interest (ROI) containing the pixels which are used in fitting the tracked particle's position. The choice of ROI is particularly important for the centroid method where inclusion of background pixels increases the localization error---in the extreme case of an infinite ROI, the localization error of the centroid method is infinite.
Second, the derivation of the localization error in presence of motion blur assumes an effectively symmetrical recorded PSF.
We expect the first assumption to break down for low values of $P$ and the second to break down for high motion blur.

To confirm the analytical approach for cases where we expect it to hold, and to investigate cases where it does not, we performed Monte Carlo simulations of a point-like diffusing fluorescent particle emitting photons recorded through a microscope by a CCD or CMOS camera. From such images we used an automated procedure for selecting the ROI and fitting the PSF recorded inside this ROI in order to determine the precision of localization methods in practice.

Images were simulated using a continuous variant of the exact Gillespie algorithm~\cite{Gillespie1977}, which uses that since photon emission is a Poisson process, the times elapsed between a particle emits two consecutive photons are exponentially distributed.
The particle was started out at $(x_{\rm true},y_{\rm true})=(0,0)$.
An exponentially distributed waiting time until emission of the first photon was then drawn, $\tau_1\sim{\rm Exp}(r)$. [Here $\tau_1\sim{\rm Exp}(r)$ is short for $\tau_1$ is exponentially distributed with rate $r$.]
The displacement undergone by the particle in each perpendicular direction during the time-interval $\tau_1$ was then drawn as $dx_{\rm true},dy_{\rm true}\sim N(0,2D\tau_1)$, i.e., both normally distributed with mean zero and variance $2D\tau_1$.
The position of the particle at time $t=\tau_1$ was then $(x_{\rm true},y_{\rm true})=(dx_{\rm true},dy_{\rm true})$; from this position the particle emitted a photon whose apparent position, as recorded by the camera, was equal to the particle's true position plus a photon noise term due to diffraction in the microscope, $\xi_x,\xi_y\sim N(0,s_a^2)$.
A new waiting time $\tau_2\sim{\rm Exp}(r)$ was drawn; the particle's position was updated by adding $dx_{\rm true},dy_{\rm true}\sim N(0,2D\tau_2)$ to $(x_{\rm true},y_{\rm true})$; the particle emitted a photon from its new position which was recorded with a photon noise term, $\xi_x,\xi_y\sim N(0,s_a^2)$.
The procedure was repeated until $\sum_{i=1}^{P+1}\tau_i>\dt$, where the last photon (corresponding to $P+1$) was not recorded.
The recorded photon positions were then compared to a 64$\times$64 pixel grid of individual dimensions $a\times a=100~\nm\times100~\nm$  (the grid was large enough that the particles did not diffuse out of the ``camera" during the time-lapse); each position falling inside a given pixel added one to its count.
Finally, Poisson distributed background noise was added to each pixel with mean $b^2=qPa^2/(2\pi s_a^2)$.

The resulting image, $I$, was then treated to estimate the particle's average position.
A thresholding procedure was performed which removed all pixels under a certain threshold equal to $n_{\rm thres}b^2$, yielding a binary matrix $A$, with ones in pixels where the photon count was above the threshold and zeros where it was below.
To remove single background pixels that were over the threshold due to random fluctuations, binary erosion of $A$ by a $3\times3$ matrix was performed.
The ROI was then expanded by a number $n_{\rm dilate}$ of successive binary dilations of $A$ by a $3\times3$ matrix.
The thresholds $n_{\rm thres}$ and $n_{\rm dilate}$ were chosen for the highest localization precision, and depended on $P$; for GME and MLEwG, $n_{\rm dilate}$ needed only be large enough to include a substantial part of the PSF, while for the centroid method, $n_{\rm dilate}$ needed to be chosen as function of $n_{\rm thres}$ and $P$ to maximize precision.
The particle was then localized~\cite{Mortensen2010,Deschout2012} using only pixels of $I$ that corresponded to non-zero entries of $A$. For the centroid method, fitting involved first subtracting the average background amplitude from all pixels~\cite{Deschout2012}; the average background was estimated from pixels in a perimeter of three pixels around the ROI.

The above procedure was repeated 1\,000 times for each set of parameter values in order to estimate the localization error $\s_{\rm p}$ of the various methods and the probability $\epsilon_{\rm p}$ for localization to fail in practice.

For the case where particles are assumed to be successfully localized in all images, the values $\s_{\rm p}$ were used in Eqs.~(\ref{eq:info2}) and (\ref{eq:var(D_1)}) to obtain numerical numerical estimates of the CRB.
For cases where localization fails for some images, the numerical estimates of the CRB were found using both $\s_{\rm p}$ and $\epsilon_{\rm p}$ as described in Section~\ref{sec:optimal}.

\section{Supplemental figures}
\label{app:supFigs}
\begin{figure*}
  \centering
  \includegraphics{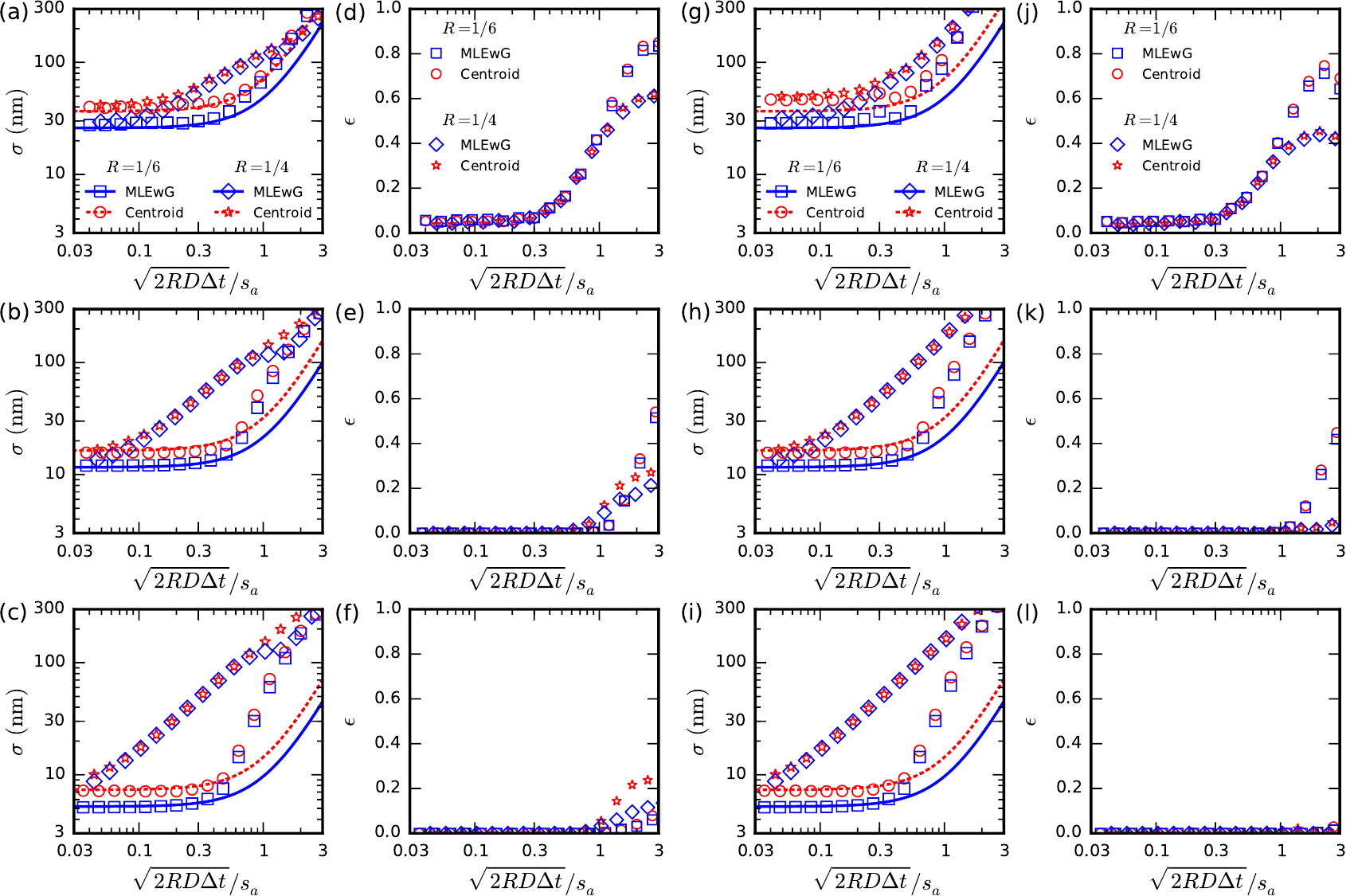}
  \caption{(Color online) Performance of localization methods as function of normalized motion blur, $\sqrt{2RD\dt}/s_a$, where localization is considered to have failed if the error between the estimated and true average positions is higher than (a)--(f) $2s=2\sqrt{s_a^2+2RD\dt}$ or (g)--(l) $4s=4\sqrt{s_a^2+2RD\dt}$.
  (a)--(c),(g)--(i) Amplitude of localization errors, $\s$, and (d)--(f),(j)--(l) probability of localization to fail.
  Lines mark theoretical errors [Eqs.~(11) and (12)] and symbols mark mean errors (error bars the values are smaller than symbol sizes) averaged over 1\,000 MC simulations.
  The number signal photons per image are (a),(d) 200, (b),(e) 1\,000, and (c),(f) 5\,000.
  The width of the stationary PSF is $s_a=153\ \mathrm{nm}$, the background-to-signal ratio is $q=1$, and the results are for 2D diffusion in the image plane.}
  \label{supfig:sig(P)}
\end{figure*}

\begin{figure*}
  \centering
  \includegraphics{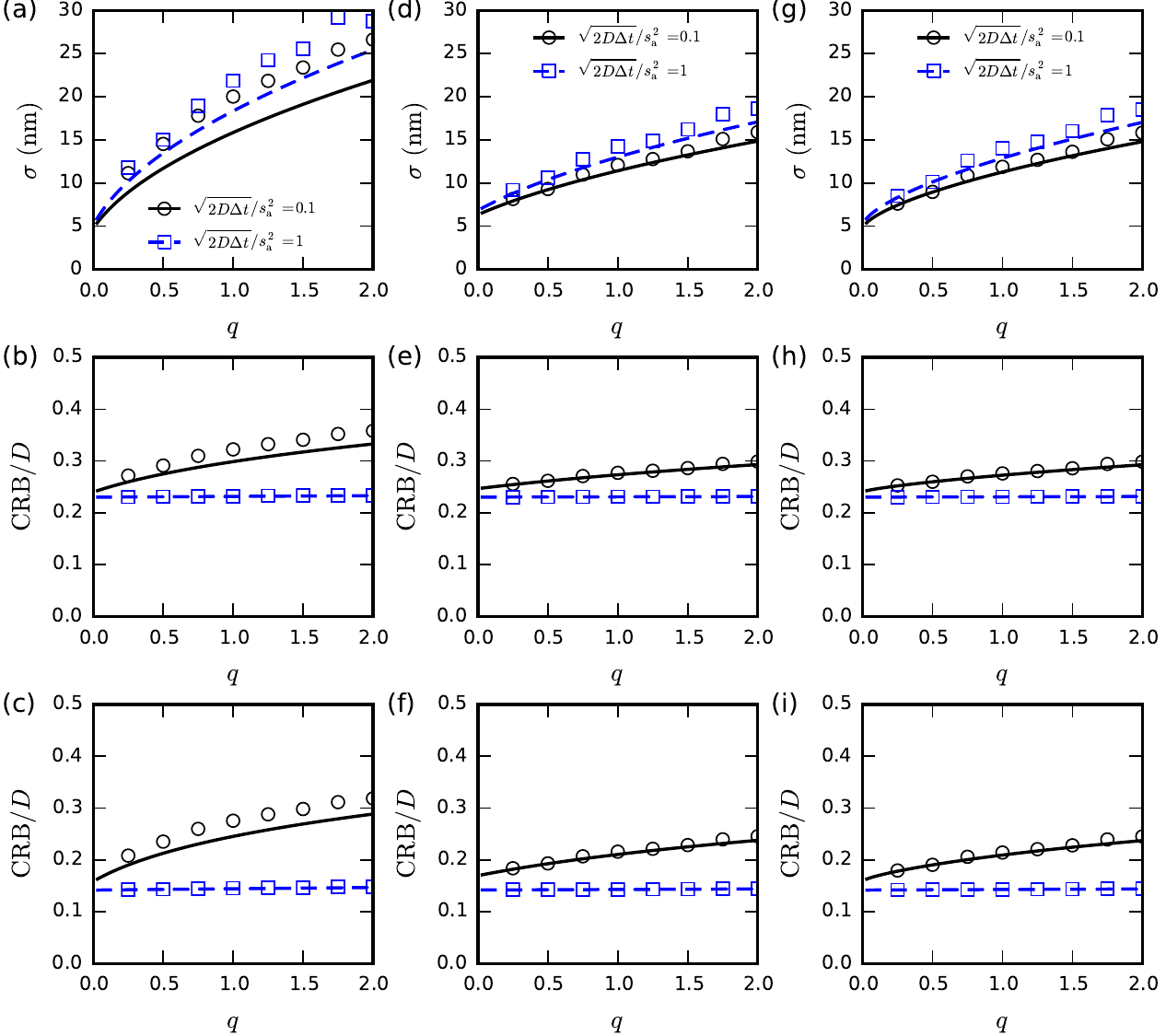}
  \caption{(Color online) Influence on background noise on localization error and estimator precision for continuously open shutter ($R=1/6$).
  (a),(d),(g) Localization error as function of the background-to-signal ratio $q$.
  (b),(c),(e),(f),(h),(i) Cram\'er-Rao bound (CRB) on the standard error of any unbiased estimator of diffusion coefficients as a function of $q$ for: (b),(e),(h) unknown localization error variance $\s^2$ and (c),(f),(i) known $\s^2$;
  Results are shown for (a)--(c) centroid, (d)--(f) GME, and (g)--(i) MLEwG localization.
  Increasing $q$, i.e., increasing the background noise, adversely affects the precision of the centroid method 
  , though only for low $\sqrt{2D\dt}/s_a$; for low background the precision of the centroid method approaches that of MLEwG.
  The precision of GME is lower than MLEwG for low $q$, though rapidly approaches it as $q$ is increased.
  Changing $q$ does however not change the qualitative results presented in Figs.~6--9.
  In all plots, the number of photons recorded per image is $P=1\ 000$, the time-series length is $N=100$, the stationary PSF width is $s_a=153\ \mathrm{nm}$, and the particle diffuses in 2D.}
  \label{fig:CR(b)}
\end{figure*}

\begin{figure*}
  \includegraphics{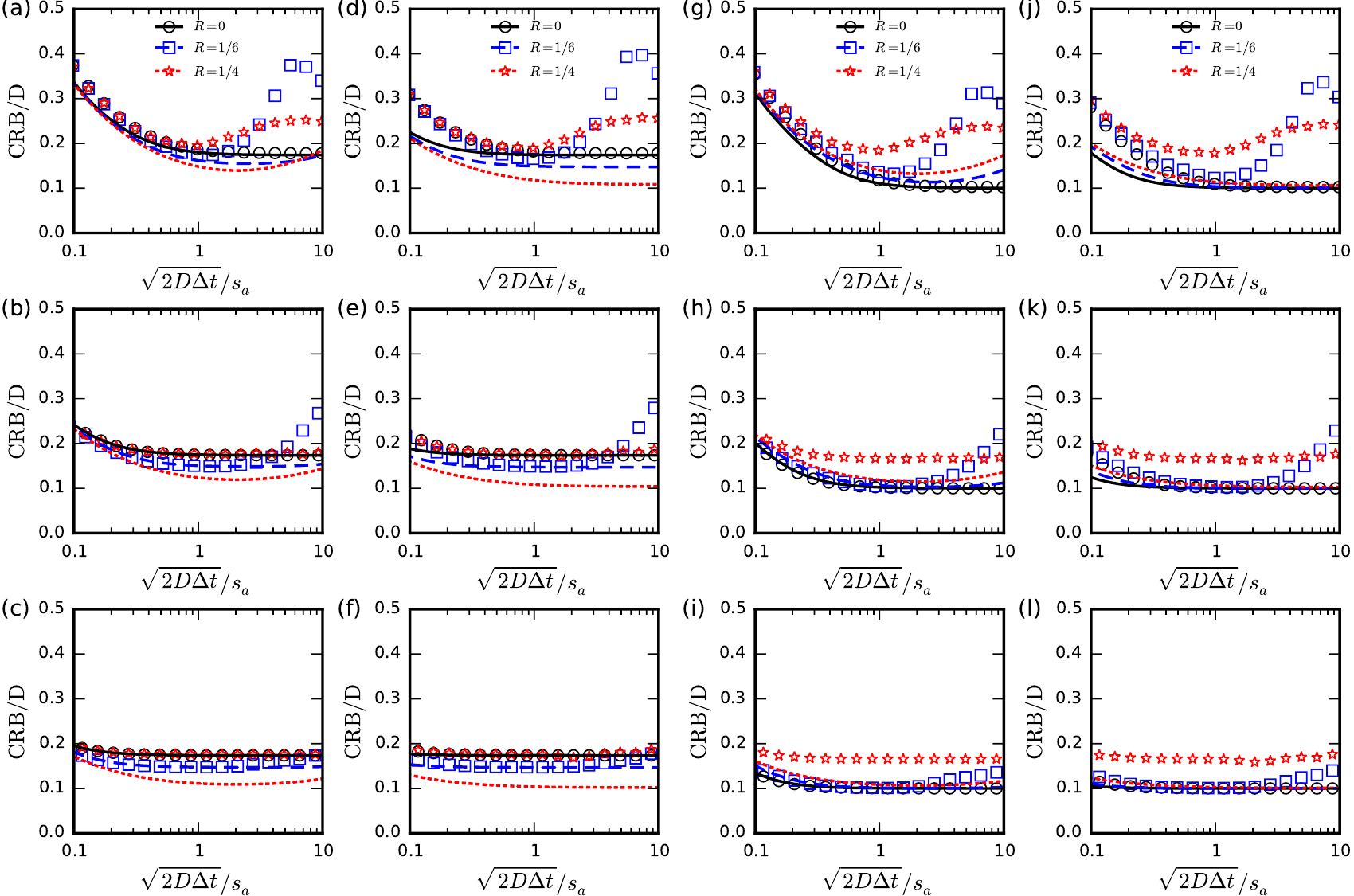}
  \caption{(Color online) Cram\'{e}r-Rao bound (CRB) on the standard error of any unbiased estimator of the diffusion coefficient in the presence of motion blur where a fraction $\epsilon$ of the particle's positions are missing in the recorded time-series: (a)--(f) for unknown $\s$, (g)--(l) for known $\s$.
  The centroid method is used for localization in (a)--(c) and (g)--(i), while MLEwG is used in (d)--(f) and (j)--(l).
  The number of signal photons recorded per image is in (a),(d),(g),(j) 200, (b),(e),(h),(k) 1\,000, and (c),(f),(i),(l) 5\,000.
  In all panels results are shown for 2D diffusion, the background-to-signal ratio is $q = 1$, the number of recorded images is $N+1 = 101$, and the time-series length is $(1-\epsilon)N$ [see Figs.~\ref{fig:sig}(d)--\ref{fig:sig}(f) for corresponding values of $\epsilon$].}
  \label{supfig:sig}
\end{figure*}

\begin{figure*}
  \includegraphics{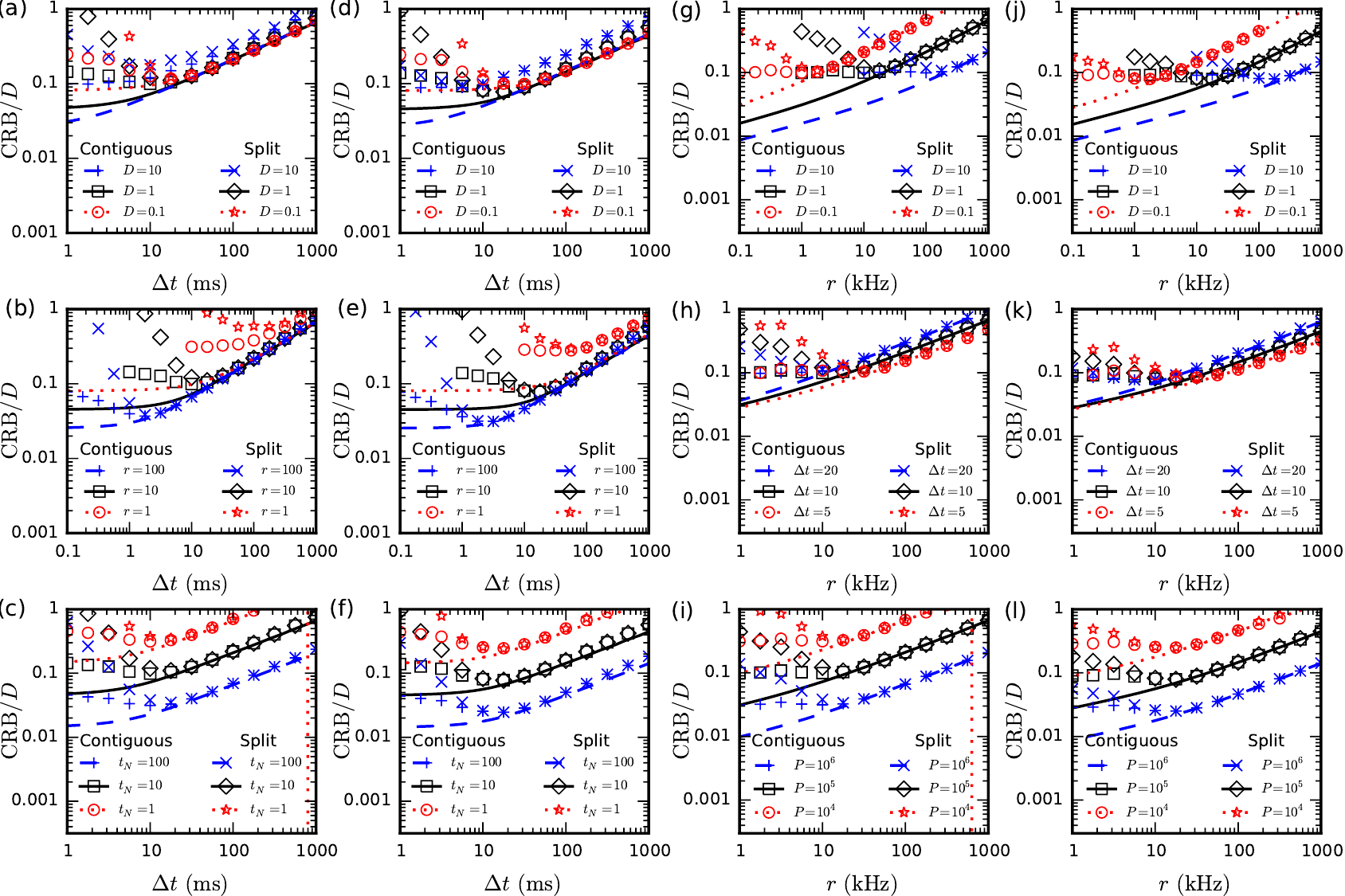}
  \caption{(Color online) Influence of the value of physical parameters on the normalized Cram\'er-Rao bound on the standard error of any unbiased estimator of the diffusion coefficient (${\rm CRB}/D$): (a)--(f) for limited experimental recording time; (g)--(l) for limited photostability of the fluorescent marker.
  (a),(d),(g),(j) Influence of the value of the diffusion coefficient $D$ (values for $D$ given in legends are in units of $\mu{\rm m}^2{\rm s}^{-1}$).
  (b),(e) Influence of the value of the photon emission rate $r$ (values in legends are given in kHz).
  (c),(f) Influence of the limit $t_N$ on the length of the recorded time-series (values in legends are given in s).
  (h),(k) Influence of the time-lapse $\dt$ (values in legends are given in ms).
  (i),(l) Influence of the total number of photons recorded, $P$.
  In all panels, results are for 2D diffusion, background-to-signal ratio $q=1$, and stationary PSF width $s_a=153\ {\rm nm}$.}
  \label{supfig:parameters}
\end{figure*}

This appendix contains four supplemental figures that support the conclusions and discussion in the main text.
Figure~\ref{supfig:sig(P)} shows that localization error and probability of failure only depend weakly on the choice of the threshold that defines when localization is considered to have failed.
Figure~\ref{fig:CR(b)} shows how localization error and CRB depend on background noise (i.e. the value of the background-to-signal ratio $q$).
Figure~\ref{supfig:sig} shows how the CRB depends on $\sqrt{2D\dt}/s_a$ when the effect of localization failure is taken into account.
Figure~\ref{supfig:parameters} investigates how the values of physical parameters, i.e., $D$, $r$, $t_N$, $\dt$, and $P_N$, influence the optimal choice of experimental parameters.

\begin{thebibliography}{10}
\bibitem{note-signalPhotons}
\bibinfo{note}{Note that $P$ differs from the total number of photons emitted from the fluorophore and recorded by the camera, which includes also photons in the power-law tails of the PSF; $P$ is roughly $40\%$ smaller than this number~\cite{Mortensen2010}.}
\bibitem{note-fail}
\bibinfo{note}{Note that while the choice of the threshold is arbitrary, its specific value does not influence the results qualitatively (see Appendix~\ref{app:supFigs}).}
\bibitem{note-blur}
\bibinfo{note}{Note that the theoretical limits $R=0$ and $R=1/4$, corresponding to an illumation function given by one or two delta-functions, respectively, cannot be obtained in practice since the camera integration time must necessarily be finite. Similarly, the read-out process of the camera may result in a non-zero time-lag between images and, consequently, a motion blur coefficient that is slightly smaller than $1/6$ when recording with continuously open shutter.
These limits are useful theoretical abstractions, however, as they limit the behavior that may be observed; the slight deviations that may be expected in practice do not change our conclusions since we are concerned with the qualitative goal of choosing the optimal strategy for experimental design.}
\bibitem{note-P(R)}
\bibinfo{note}{For simplicity, we also assume here that the signal and background photon counts $P$ and $b^2$, respectively, are unchanged, i.e., that one may record as many photons in a frame using a stroboscopic technique as by leaving the shutter continuously open; we discuss below the validity of this assumption and how it affects the results.}
\end{thebibliography}
\end{document}